\documentclass[aps,prb,twocolumn,superscriptaddress]{revtex4-1}

\usepackage{graphicx}
\usepackage{amsmath}

\begin{document}


\title{Electrical manipulation of a ferromagnet by an antiferromagnet}


\author{V.~Tshitoyan}
\author{C.~Ciccarelli}
\affiliation{Microelectronics Group, Cavendish Laboratory, University of Cambridge, CB3 0HE, UK}
\author{A.~P.~Mihai}
\thanks{Currently at London Centre for Nanotechnology, Department of Materials, Imperial College London, SW7 2BP, UK}
\author{M.~Ali}
\affiliation{School of Physics and Astronomy, University of Leeds, Leeds LS2 9JT, UK}
\author{A.~C.~Irvine}
\affiliation{Microelectronics Group, Cavendish Laboratory, University of Cambridge, CB3 0HE, UK}
\author{T.~A.~Moore}
\affiliation{School of Physics and Astronomy, University of Leeds, Leeds LS2 9JT, UK}
\author{T.~Jungwirth}
\affiliation{Institute of Physics ASCR, v.v.i., Cukrovarnick\'a 10, 162 53 Praha 6, Czech Republic}
\affiliation{School of Physics and Astronomy, University of Nottingham, Nottingham NG7 2RD, UK}
\author{A.~J.~Ferguson}
\email[]{ajf1006@cam.ac.uk}
\affiliation{Microelectronics Group, Cavendish Laboratory, University of Cambridge, CB3 0HE, UK}


\date{\today}

\begin{abstract}
We demonstrate that an antiferromagnet can be employed for a highly efficient electrical manipulation of a ferromagnet. In our study we use an electrical detection technique of the ferromagnetic resonance driven by an in-plane ac-current in a NiFe/IrMn  bilayer. At room temperature, we observe antidamping-like spin torque acting on the NiFe ferromagnet, generated by the in-plane current driven through the IrMn antiferromagnet.  A large enhancement of the torque, characterized by an effective spin-Hall angle exceeding  most heavy transition metals, correlates with the presence of the exchange-bias field at the NiFe/IrMn interface. It highlights that, in addition to strong spin-orbit coupling, the antiferromagnetic order in IrMn governs the observed phenomenon.
\end{abstract}


\maketitle

\section{\label{sec:introduction}Introduction}

Recently, a new direction in spintronics has been proposed based on non-relativistic~\cite{Nunez2006,Haney2008,Xu2008,Gomonay2010,Hals2011} and relativistic~\cite{Shick2010,Zelezny2014} spin-transport phenomena in which antiferromagnets (AFMs) complement or replace ferromagnets (FMs) in active parts of the device. AFMs have for decades played a passive role in conventional spin-valve structures where they provide pinning of the reference FM layer~\cite{Nogues1999}.  This implies that on one hand, incorporation of some AFM materials, including IrMn, in common spintronic structures is well established. On the other hand, limiting their utility to a passive pinning role leaves a broad range of spintronic phenomena and functionalities based on AFMs virtually unexplored. In addition to the insensitivity to magnetic fields and the lack of stray fields, AFMs are common among metals, semiconductors, and insulators and can have orders of magnitude shorter spin-dynamics timescales, to name a few immediate merits of the foreseen concept of AFM spintronics. 
  
AFM magneto-resistor and memory functionalities have been demonstrated by manipulation of the AFM moments via a FM sensitive to external magnetic fields~\cite{Park2011b,Wang2012a,Marti2014,Fina2014}. Wadley~{\em et al.}~\cite{Wadley2015} showed that in AFMs with specific crystal and magnetic structures AFM moments can be manipulated electrically. Several studies have also focused on transmission and detection of spin-currents in AFMs. In FM/AFM/normal-metal (NM) trilayers, a spin-current was pumped from the FM, detected by the inverse spin-Hall effect (ISHE) in the NM, and the observed robust spin-transport through the interfacial AFM (insulating NiO) was ascribed to AFM moment fluctuations~\cite{Wang2014d,Hahn2014}. Efficient spin transmission through an AFM (IrMn) was also inferred from an inverse experiment in the FM/AFM/NM structure~\cite{Moriyama} in which spin-current was generated by the spin-Hall effect (SHE) in the NM and absorbed via the spin transfer torque (STT)~\cite{Ralph2008} in the FM. Measurements in FM/AFM bilayers have demonstrated that a metallic AFM itself (e.g. IrMn) can act as an efficient ISHE  detector of the spin-current injected from the FM, with comparable spin-Hall angles to heavy NMs~\cite{Mendes2014,Zhang2014}.

Our work makes the next step beyond previous studies of transmission and detection of spin-currents in AFMs by focusing on spin manipulation by AFMs. In a NiFe/Cu/IrMn structure we demonstrate that the IrMn AFM produces a large SHE spin-current which is transmitted through Cu and exerts an antidamping-like  STT on the NiFe FM comparable in strength to the SHE-STT generated by Pt. Upon removing the interfacial Cu layer, we observe that the size of the antidamping-like torque is strongly enhanced and that it correlates with the exchange-bias field associated with the fixed AFM moments at the coupled NiFe/IrMn interface. Our observations point to new physics and functionalities that AFMs can bring to the currently highly active research area of relativistic spin-orbit torques induced by in-plane currents in inversion asymmetric magnetic structures~\cite{Bernevig2005c,Manchon2008,Chernyshov2009,Fang2011,Miron2010,Miron2011,Liu2012,Garello2013,Kurebayashi2014}.
\begin{figure}[ht]
	\includegraphics[width=.95\columnwidth,angle=0]{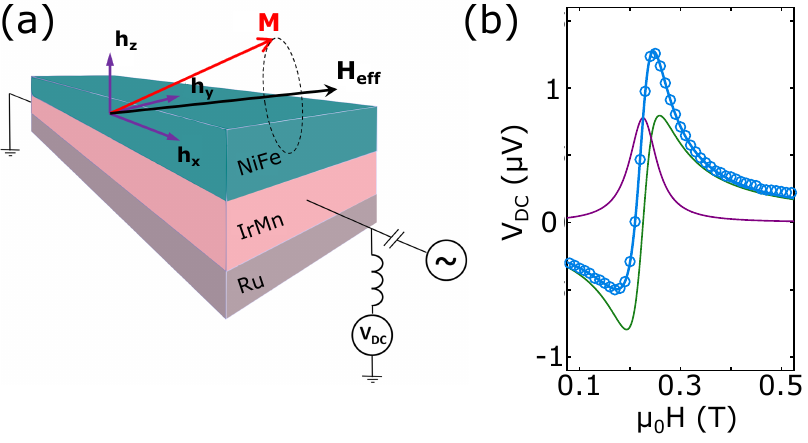}
	\caption{\textbf{Spin-orbit FMR experiment.} (a) Schematic representation of the measurement technique. MW current-induced effective field $\textbf{h}(h_x,h_y,h_z)$ drives magnetization precession around the total field $\mathbf{H_{eff}}$. Precessing magnetization results in oscillating resistance due to AMR. This mixes with oscillating current of the same frequency resulting in a measurable DC voltage. (b) Resonance curve decomposed into symmetric and antisymmetric components measured in a bar with 2~nm IrMn at frequency of 17.9 GHz.}
	\label{fig:figure1}
\end{figure}
\section{\label{sec:measurements}Measurements}
\begin{figure}[b]
	\includegraphics[width=.95\columnwidth,angle=0]{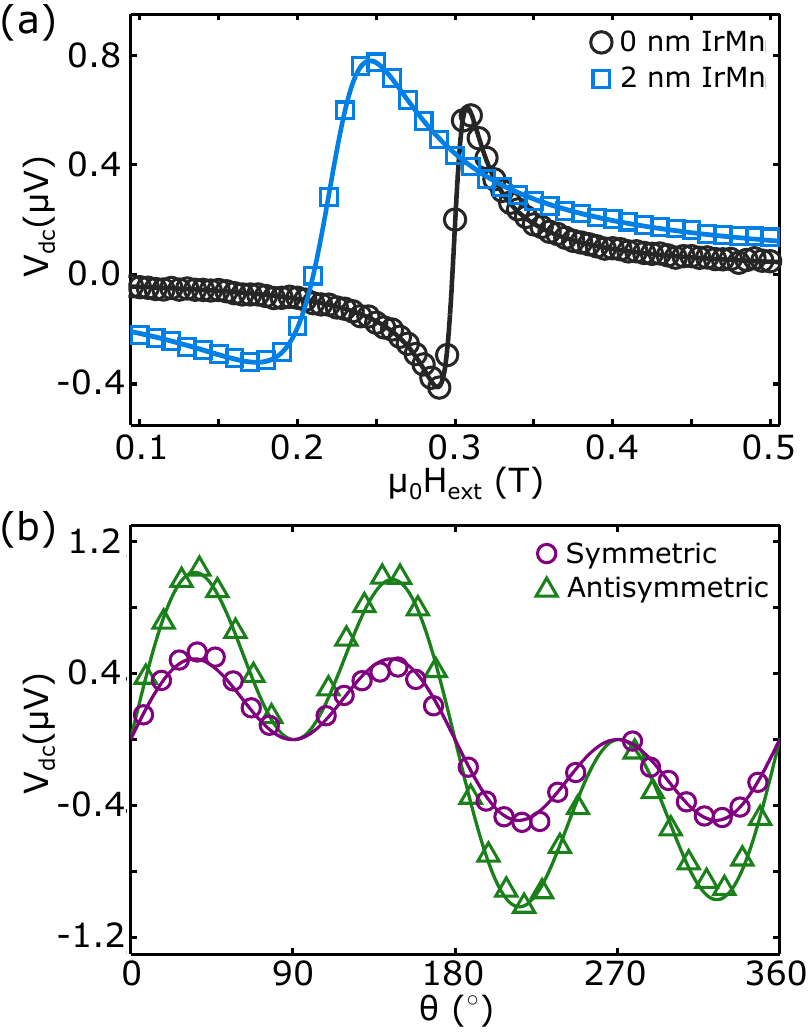}
	\caption{\textbf{AFM-induced torque and its symmetries.} (a) Comparison of resonance curves measured in samples with and without the IrMn layer. Both measurements are performed at 17.9 GHz, $\theta = 45^\circ$. Antisymmetric components are normalized to 1~$\mathrm{\mu{V}}$. (b) Symmetries of $V_{sym}$ and $V_{asy}$ for the sample with 2~nm IrMn. Solid lines are fits to equations \ref{eqn:sym} and \ref{eqn:asy}.}
	\label{fig:resonance-comparison}
\end{figure}
\begin{figure*}[t]
	\includegraphics[width=2.00\columnwidth,angle=0]{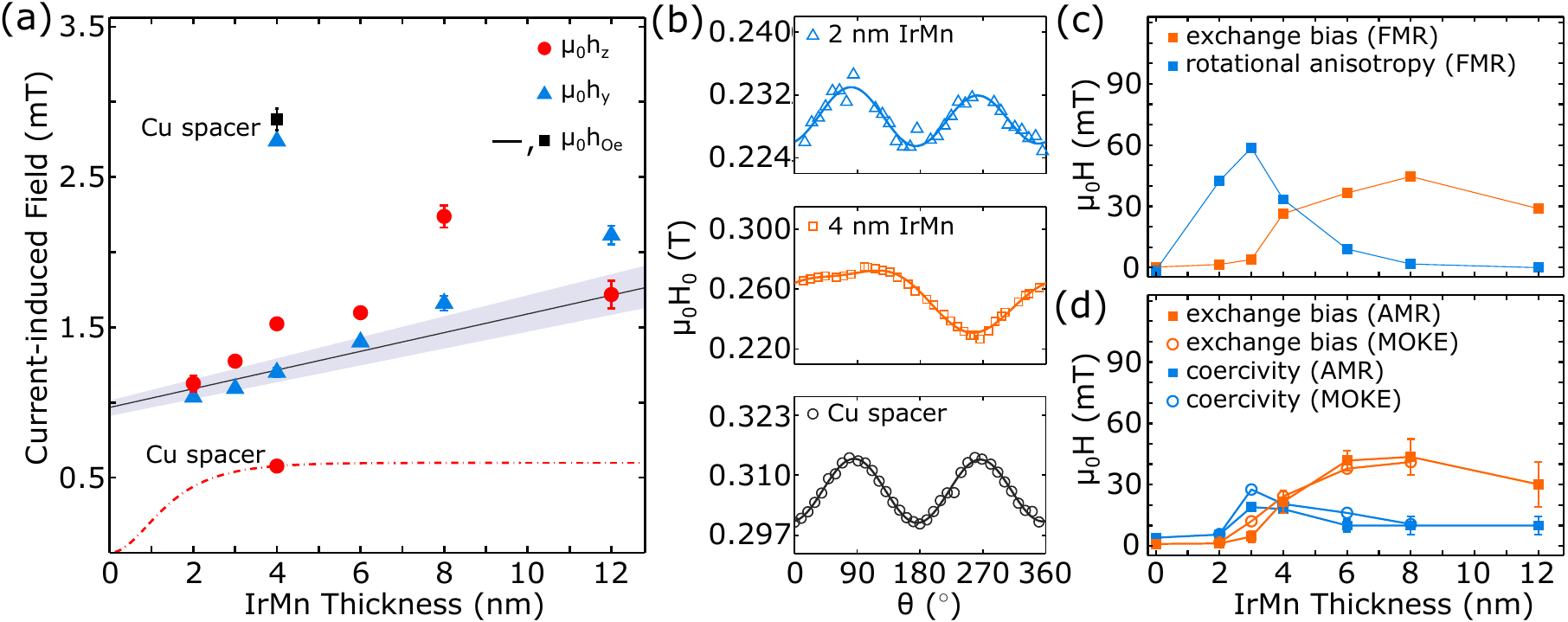}
	\caption{\textbf{AFM thickness dependence of current-induced fields and anisotropies.} (a) $h_z$, $h_y$ and calculated Oersted field $h_{Oe}$ for 1.8~$\mathrm{\mu{m}}$ wide bars with different IrMn thicknesses, as well as the sample with the 2~nm Cu spacer layer. The results are normalized to a current density of $10^7$~$\mathrm{A/cm^2}$ in IrMn. The shaded area around $h_{Oe}$ is the error due to uncertainties in layer resistivities, whereas the error bars of $h_z$ and $h_y$ are due to the standard errors from the fitting of the symmetries, AMR and MW current. The systematic uncertainties in layer resistivities have not been included in the error bars of $h_z$ and $h_y$, however this uncertainty, which is approximately 20 \%, is included in the values of effective spin-Hall angles in the main text. The dotted line is the estimated spin-Hall effect contribution to $h_z$ for $\lambda_{sd} = 1$~nm. (b) Angle dependences of resonance field for the samples with 2 and 4~nm IrMn thicknesses, as well as the sample with the 2~nm Cu spacer layer. Solid lines are fits taking into account unidirectional, uniaxial and rotational anisotropies. (c) IrMn thickness dependence of the exchange bias and the rotational anisotropy extracted from the fits in (a). (d) IrMn thickness dependence of exchange bias and coercivity extracted from hysteresis loops measured using MOKE and AMR switching.}
	\label{fig:figure3}
\end{figure*}
\begin{figure}[ht]
	\includegraphics[width=.95\columnwidth,angle=0]{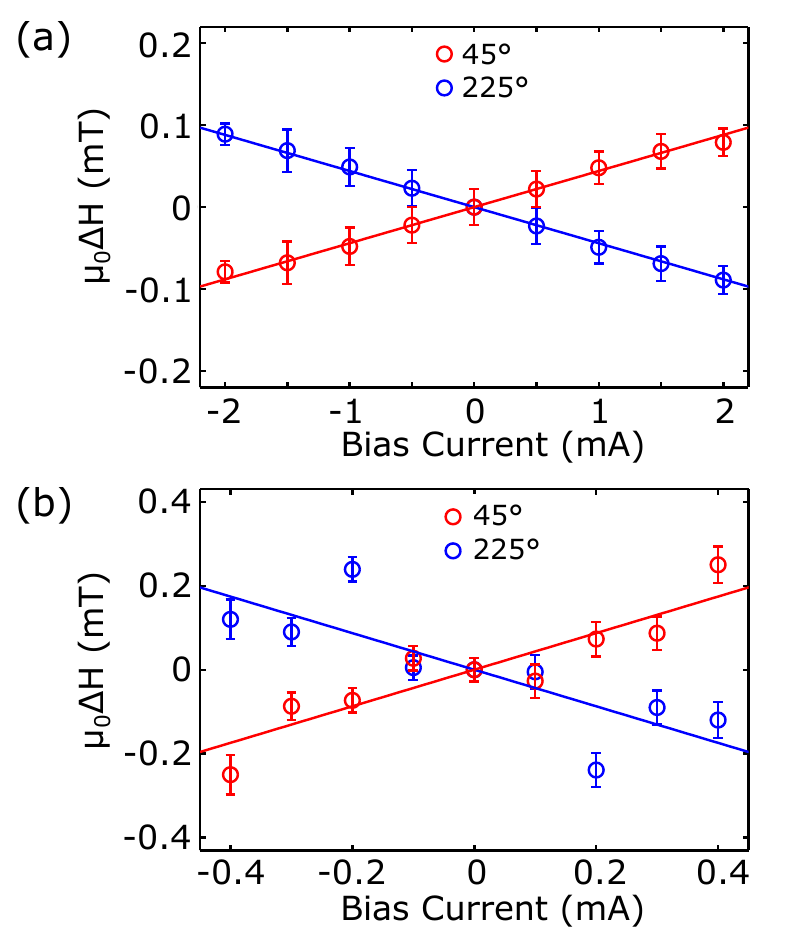}
	\caption{\textbf{DC bias dependence of the FMR linewidth.} (a) Change of FMR linewidth with DC current for the IrMn(4)/Cu(2)/NiFe(4) structure measured at $\omega/2\pi = 8$ GHz and (b) IrMn(2)/NiFe(4) structure measured at $\omega/2\pi = 14.1$ GHz, for two different directions of magnetization with respect to the current. The data points are extracted using the linewidth difference between positive and negative bias currents.}
	\label{fig:figure4}
\end{figure}
Multilayers $\mathrm{SiO_{x}/Ru(3)/IrMn(d_A)/NiFe(4)/Al(2)}$ and  $\mathrm{SiO_{x}/Ru(3)/IrMn(4)/Cu(d_N)/NiFe(4)/Al(2)}$ used in our measurements were grown using dc magnetron sputtering. The numbers represent layer thicknesses in nm, IrMn thickness $d_A$ in the first type of multilayers varies from 0 - 12~nm, and Cu thickness $d_N$ in the second type of multilayers is 1 or 2~nm. We apply microwave (MW) frequency electrical current to a bar patterned from the magnetic multilayer. Bars used in our measurements vary from 500~nm to 4~$\mathrm{\mu{m}}$ in width and 5~$\mathrm{\mu{m}}$ to 240~$\mathrm{\mu{m}}$ in length. Torques induced by the oscillating current in the bar drive magnetization precession of the NiFe around the equilibrium axis defined by an applied saturating magnetic field. A diagram of the measurement setup and the device is shown in Fig.~\ref{fig:figure1}(a). The bar is aligned along the x-axis, while the z-axis represents the out-of-plane direction. Resonant precession is detected as a rectified dc voltage due to anisotropic magnetoresistance (AMR)~\cite{Costache2006}. In our studies we keep the frequency of the current constant and sweep the in-plane magnetic field (Fig.~\ref{fig:figure1}(b)).

From the decomposition of the resonance into symmetric and antisymmetric Lorentzians~\cite{Fang2011} we deduce the out-of-plane and in-plane components of the driving field as
\begin{equation}
V_{sym} = \frac{I\Delta{R}}{2}A_{sym}h_{z}sin{2\theta}
\label{eqn:sym}
\end{equation}
\begin{equation}
V_{asy} = \frac{I\Delta{R}}{2}A_{asy}(h_{y}\cos\theta - h_{x}\sin\theta)\sin{2\theta}.
\label{eqn:asy}
\end{equation}
Here $I$ is the current in the bar, $\Delta{R}$ is the AMR amplitude, $A_{sym}$ and $A_{asy}$ are coefficients determined by the magnetic anisotropies, and $\theta$ is the angle between the magnetization and current directions. Current-induced fields $h_x, h_y$ and $h_z$ can be obtained from the measured angle-dependences of $V_{sym}$ and $V_{asy}$. We calibrate the microwave current $I$ in the bar from the resistance change induced by microwave heating (Supplementary Section S1). $\Delta{R}$ is obtained from the in-plane AMR measurement using a 1~T magnetic field, while the anisotropy coefficients $A_{sym}$ and $A_{asy}$ are extracted from the angle dependence of the resonance field (Supplementary Section S4).

In Fig.~\ref{fig:resonance-comparison}(a) we compare resonance curves for samples without the Cu layer and with 0 and 2~nm thick IrMn. The resonance is predominantly antisymmetric without IrMn, indicating a driving field in the in-plane direction. The resonance then acquires a substantial symmetric component in the presence of the AFM, indicating an additional driving field in the out-of-plane direction. Both symmetric and antisymmetric components follow a $\sin{2\theta}\cos{\theta}$ angle dependence (Fig.~\ref{fig:resonance-comparison}(b)). This means that the in-plane effective field is along the y direction and is independent on the magnetization direction, resulting in an out-of-plane field-like torque, $\mathbf{\tau_{z}} \propto \mathbf{m} \times \mathbf{\hat{y}}$. In contrast, $h_{z}$ depends on magnetization direction as $\cos\theta \propto [\mathbf{j} \times \mathbf{\hat{z}}] \times \mathbf{m}$, thus resulting in an antidamping-like in-plane torque $\mathbf{\tau_{ad}} \propto \mathbf{m} \times ([\mathbf{j} \times \mathbf{\hat{z}}] \times \mathbf{m})$.

We find that for all our samples the magnitude of $h_y$ is compatible with the magnitude of the Oersted field induced by the current in IrMn and Ru layers. The Oersted field is calculated using the individual layer resistivities extracted from resistance measurements of bars with different IrMn and Ru thicknesses, as described in Supplementary Sections S2 and S3 ($\rho_{IrMn} = 20.5 \pm 3.3 \times 10^{-7}$ $\mathrm{\Omega{m}}$, $\rho_{Ru} = 4.0 \pm 0.3 \times 10^{-7}$ $\mathrm{\Omega{m}}$, and $\rho_{NiFe} = 5.4 \pm 0.4 \times 10^{-7}$ $\mathrm{\Omega{m}}$). From the fits of the symmetric and antisymmetric components to Eqs.~(\ref{eqn:sym}) and (\ref{eqn:asy}) shown in Fig.~\ref{fig:resonance-comparison}(b) we deduce $\mu_0h_z = 1.13 \pm 0.05$~$\mathrm{mT}$ and $\mu_0h_y = 1.04 \pm 0.03$~$\mathrm{mT}$, while for the Oersted field we find $\mu_0h_{Oe} = 1.09 \pm 0.07$~$\mathrm{mT}$. All values reported for the current-induced fields are normalised to a current density of  $10^7$~$\mathrm{A/cm^2}$ in IrMn.

The symmetry of $h_z$ is compatible both with the antidamping-like term of the interface-induced Rashba spin-orbit torque~\cite{Miron2010}, as well as with the SHE-STT~\cite{Miron2011,Liu2012}. In the latter case the spin-current generated in the IrMn by the SHE drives magnetization precession in the NiFe layer by STT. Both of these effects occur in FM/NM structures, however, we show  that additional effects arise due to the AFM nature of IrMn and the exchange coupling at the FM/AFM interface. 

To separate the contribution of the exchange-coupled NiFe/IrMn from the SHE-STT, we performed measurements in samples with 4~nm thick IrMn, and 1 and 2~nm thick Cu spacers between IrMn and NiFe. Cu has a spin-diffusion length of 350~nm~\cite{Yakata2006} and thus 2~nm of Cu would transfer $>$99\% of the spin-Hall current from IrMn, but eliminate the FM/AFM coupling and the FM/AFM interface-induced effects. 

Results obtained in samples with the Cu spacer and without Cu and  different IrMn thicknesses  are summarized in Fig.~\ref{fig:figure3}(a). Firstly, one can see that the $h_z$ field does not vanish with the introduction of Cu, indicating the SHE in IrMn. From the value of $h_z$ we can obtain the spin-Hall angle $\theta_{SH}$ of IrMn from the expression
\begin{equation}
\theta_{SH} = \frac{2e{\mu_0M_s}{d_F}}{\hbar{J_{IrMn}}}{h_z}.
\end{equation}
Here $d_F = 4$ $\mathrm{nm}$ is the thickness of the NiFe layer, $\mu_0M_s = 1$ $\mathrm{T}$ is the saturation magnetization of NiFe, $J_{IrMn} = 10^7$~$\mathrm{A/cm^2}$ is the charge current density in IrMn and $\mu_0h_z = 0.58 \pm 0.02$~$\mathrm{mT}$ is obtained from the measurement. We get $\theta_{IrMn} = 0.056 \pm 0.009$, in good agreement with the expected value for $\mathrm{Ir_{25}Mn_{75}}$~\cite{Mendes2014}. Here the uncertainty also includes the uncertainty of the current density in IrMn from the layer resistivity calibration. It is important to mention that the same value of $\theta_{IrMn}$ was obtained for both 1~nm and 2~nm Cu spacers, as well as bars with 1.8~$\mathrm{\mu{m}}$ and 500~nm widths. Remarkably, in addition to the SHE, we see a large contribution from the FM/AFM interface in samples without Cu, initially increasing with the IrMn thickness and with a peak at 8~nm of IrMn, with a magnitude corresponding to an effective spin-Hall angle of $0.22 \pm 0.04$. The values of effective spin-Hall angles for two samples, as well as the damping-like nature of $h_z$ were confirmed by measuring the dc bias dependence of the FMR linewidth \cite{Liu2011}. Depending on the direction of DC current with respect to FM magnetizaion, an additional damping or antidamping is induced, thereby increasing or decreasing the FMR linewidth. For the sample with the Cu spacer we obtain $\theta_{SH} = 0.043 \pm 0.001$ (Fig.~\ref{fig:figure4}(a)) and for the sample with 2 nm IrMn we get $\theta_{SH} = 0.135 \pm 0.022$ (Fig.~\ref{fig:figure4}(b)). We use

\begin{equation}
\theta_{SH} = \frac{\partial{(\mu_0\Delta{H})}}{\partial(j_{IrMn})}\times\frac{\gamma}{\omega}\frac{2e}{\hbar} \frac{(H_{res} + M_{eff}/2) \mu_0M_s t_{NiFe}}{sin\theta}
\end{equation}
where the first term is the slope of the linear fit with respect to the current density in IrMn. For comparison, the values obtained using the magnitude of $h_z$ extracted from our FMR measurements (Fig.~\ref{fig:figure3}(a)) are $0.056 \pm 0.001$ for the sample with the Cu spacer and $0.109 \pm 0.005$ for the 2 nm IrMn sample. The values are in a good agreement if we also include the resistivity calibration error of approximately 20 \% in addition to the uncertainties from the fitting.
\begin{figure}[b]
	\includegraphics[width=.95\columnwidth,angle=0]{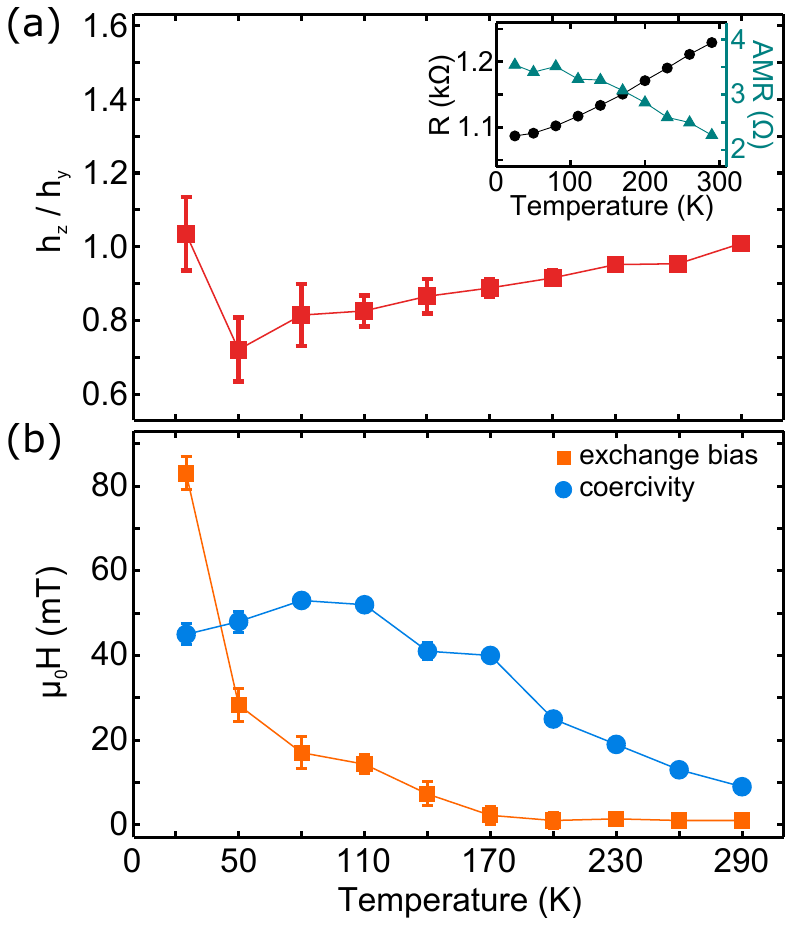}
	\caption{\textbf{Temperature dependence of current-induced fields and anisotropies.} (a) Temperature dependence of the $h_z/h_y$ ratio for the sample with 2~nm IrMn. The inset shows the temperature dependence of the AMR and total resistance of the bar. (b) Temperature dependence of the exchange bias and the coercivity for the same sample extracted from AMR switching measurements.}
	\label{fig:figure5}
\end{figure}
We note here that in a recent study, Moriyama~{\em et al.}~\cite{Moriyama} used similar FM/AFM/NM structures but instead of Ru they had Pt NM. Unlike our results, the introduction of the interfacial IrMn AFM in Moriyama~{\em et al.} structures always reduced the spin torque, compared to the reference FM/NM sample without the AFM. The authors concluded that in their case, the SHE in the AFM did not play a significant role and that the observed torque was due to the spin-Hall current from Pt transferred to the FM via spin-waves in the AFM. In our case, Ru has a small spin-Hall angle~\cite{Tanaka2008}, which we find from the control sample without IrMn to be $\approx 0.009$ (Supplementary Section S5). This, given the current distribution in the multilayer, would have a contribution of $h_z \approx 0.48$~$\mathrm{mT}$ in all the samples. Even if we assumed that the spin-angular momentum carried by the spin-Hall current from the Ru layer is fully transferred through IrMn, it would still be too small to explain the effect in samples with IrMn thicknesses larger than 3~nm, as seen in Fig.~\ref{fig:figure3}(a). 

Additionally, we performed measurements in samples with Ta seed layers instead of Ru, and found a large positive $h_z$ similar to the Ru samples ($h_z/h_y \approx 0.9$). Ta has a large negative spin-Hall angle and one would expect a negative or a largely suppressed $h_z$ if the seed layer had a significant contribution (see Supplementary Section S6 for the details).
\begin{figure*}[t]
	\includegraphics[width=2.00\columnwidth,angle=0]{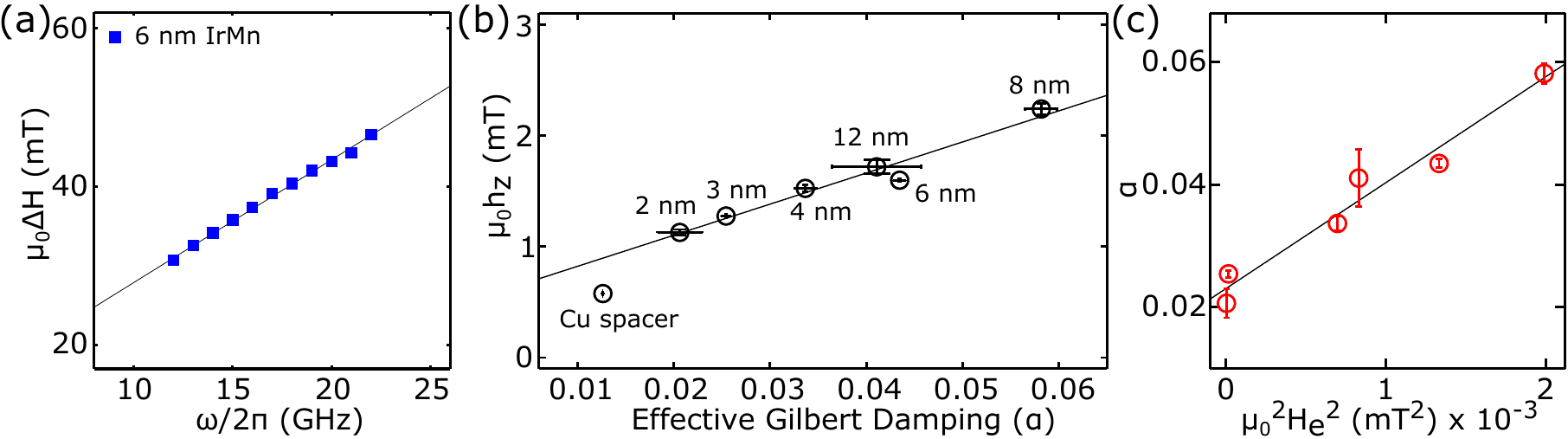}
	\caption{\textbf{AFM-induced torque and Gilbert damping.} (a) Frequency dependence of the FMR linewidth for the sample with 6 nm IrMn. The slope of the linear fit allows us to extract the effective Gilbert damping $\alpha = \gamma\mu_0(\partial\Delta{H}/\partial\omega)$, where $\gamma/2\pi$ = 28 GHz/T. (b) Current-induced out-of-plane field $h_z$ plotted against the effective Gilbert damping for samples with different IrMn thicknesses. (c) Gilbert damping is proportional to the square of the exchange bias, suggesting that one of the main damping mechanisms in our samples is the two-magnon scattering due to the inhomogeneity of the field at the FM/AFM interface induced by the exchange anisotropy.}
	\label{fig:figure6}
\end{figure*}
The increase of the antidamping-like torque in our NiFe/IrMn samples with increasing IrMn thickness cannot be explained by the increase in the spin-Hall current, as shown by the dotted line in Fig.~\ref{fig:figure3}(a), because IrMn has a spin diffusion length smaller than 1~nm~\cite{Acharyya2011, Zhang2014}. It is clearly associated with the exchange-coupled NiFe/IrMn interface. The two leading anisotropies commonly used to characterise FM/AFM interfaces are the exchange bias field and the rotational anisotropy, the latter being the origin of the increased coercivity~\cite{Stiles1999,Stiles2001}. Rotational anisotropy can be modelled as an additional effective field along the magnetization direction, and thus results in an overall decrease of the resonance field in FMR measurements. This decrease is seen in Fig.~\ref{fig:resonance-comparison}(a). The anisotropies are quantified from the angle dependence of the resonance field, plotted in Fig.~\ref{fig:figure3}(b) for the 2 and 4~nm IrMn samples and the sample with the 2~nm Cu spacer, all measured at 17.9 GHz. Comparing the top graph (2~nm IrMn with no spacer) and the bottom graph (2~nm Cu spacer), we see a smaller resonance field in the sample with 2~nm IrMn due to the rotational anisotropy induced at the FM/AFM interface, as discussed earlier. For the thicker IrMn sample (middle graph) a unidirectional contribution due to the exchange bias starts to develop. 

Thickness dependences of the exchange bias field $H_{ex}$ and the rotational anisotropy field $H_{rot}$ extracted from the fits are plotted in Fig.~\ref{fig:figure3}(c) and compared to $H_{ex}$ and $H_{c}$ extracted from MOKE and AMR switching measurements plotted in Fig.~\ref{fig:figure3}(d), showing a good agreement. One can see the onset of exchange bias at 3~nm and a peak at 8~nm of IrMn. The rotational anisotropy and coercivity are the largest for the sample with 3~nm IrMn. Similar thickness dependence has been observed experimentally using different techniques~\cite{Ali2003,McCord2004}. One can see a correlation between the size of the exchange bias and $h_z$ by comparing Fig.~\ref{fig:figure3}(a) and (c,d). It is worth mentioning here that although the exchange bias has different directions for 4 - 12~nm IrMn samples the symmetry of $h_z$ is not affected by it (Supplementary Section S7).

To confirm the correlation between the antidamping-like torque and exchange bias in one sample, we performed temperature dependence measurements of the $h_z/h_y$ ratio for the sample with 2~nm IrMn. Although this ratio is not a direct measure of the effective spin-Hall angle due to the possible current redistribution with temperature, it can help with the qualitative understanding. The results are shown in Fig.~\ref{fig:figure5}(a). The monotonous decrease in the $h_z/h_y$ ratio down to 50~K can be explained with the current redistribution in the bar. IrMn is an alloy, and thus its resistivity decreases less with temperature compared to Ru, resulting in a smaller proportion of current flowing through IrMn, and thus smaller $h_z$ at lower temperatures. The ratio can also change monotonously with temperature if there are additional temperature dependent contributions to $h_y$~\cite{Kim2014}. Nevertheless, as one can see the monotonic trend is broken below 50~K, coinciding with the abrupt increase in the exchange bias and decrease in the coercivity (Fig.~\ref{fig:figure5}(b). In the inset of Fig.~\ref{fig:figure5}(a) we plot the change of resistance and AMR with temperature, showing their monotonous behaviour for the whole temperature range. This result is significant because it shows dependence of current-induced torques on AFM-induced anisotropies in a single device. We also found that cooling down the sample from room temperature to 25 K with applied 1 T magnetic field along different directions changes the direction of the exchange bias, however, this does not significantly change magnitudes and symmetries of the current-induced fields.

\section{\label{sec:discussion}Discussion}

The origin of relativistic spin torques induced by an in-plane current at FM/NM interfaces is a subject of current intense theory discussions. Our results clearly indicate that replacing the NM with an AFM adds to the richness of these phenomena which inevitably brings more complexity to their theoretical description. To stimulate future detailed microscopic analyses we outline here possible mechanisms that might be considered as the origin of the enhancement of the antidamping-like torque and its correlation with the exchange bias. Firstly, the exchange coupling could increase the transparency at the FM/AFM interface resulting in a more efficient spin-transfer. One can estimate the efficiency of spin-transfer through FM/NM interface from the frequency dependence of the FMR linewidth~\cite{Tserkovnyak2002}. This is characterised by the effective Gilbert damping $\alpha$, extracted from the slope in Fig.~\ref{fig:figure6}(a). In Fig.~\ref{fig:figure6}(b) we plot $h_z$ as a function of $\alpha$ for the samples with different IrMn thicknesses. One can see a clear linear trend, suggesting that $h_z$ is correlated with the spin-angular momentum transfer properties through the interface. Additionally, in Fig.~\ref{fig:figure6}(c) we show that the enhancement of the spin-angular momentum transfer through the interface is indeed due to the interfacial exchange coupling, as $\alpha$ is proportional to the square of the exchange bias. This dependence also suggest that one of the main damping mechanisms in our samples is the two-magnon scattering at the FM/AFM interface, in agreement with the previous studies~\cite{Rezende2001,Yuan2009,Zhang2014}. The exact mechanism of the enhancement of $h_z$ is of complex origin due to the strong spin-orbit coupling in the system and the interface magnetic coupling. If we assume that the damping enhancement is merely due to more efficient spin-pumping and try to estimate the value of the transparency at the interface in the weak spin-orbit coupling picture of spin-mixing conductance using~\cite{Tserkovnyak2002}

\begin{equation}
G_{mix} = \frac{G_{eff}}{1-2G_{eff}\lambda_{SD}/\sigma_{IrMn}}
\end{equation}
where

\begin{equation}
G_{eff} = \frac{e^2}{h}\frac{4\pi{M_s}t_{NiFe}}{g\mu_{B}}(\alpha-\alpha_0)
\end{equation}
using values $\lambda_{SD} = 0.7$ nm~\cite{Zhang2014} and conductivity $\sigma_{IrMn} = 1/\rho_{IrMn}$, we obtain negative values for $G_{mix}$, which is non-physical. Here $\alpha_0 = 0.006$ is the Gilbert damping of bulk NiFe. One would have to assume $\lambda_{SD} < 0.1$ nm to obtain positive $G_{mix}$. This additionally suggests that the mechanism of the damping enhancement, and subsequently the torque enhancement is more complex than just an increase of spin-current transparency at the interface combined with the spin-Hall effect.

Another possibility is that additional torques are induced directly at the FM/AFM interface, or induced in the AFM and coupled to the FM via the exchange interaction. In this case the level of the magnetic order in the AFM layer could be important for the size of the torque. Wei~{\em et al.}~\cite{Wei2007} and Urazhdin~{\em et al.}~\cite{Urazhdin2007} observed changes in exchange bias in current perpendicular-to-plane geometries, attributed to torques changing the AFM magnetic structure at the FM/AFM interface. We note that our measurement is not sensitive to the bulk AFM magnetic order, except through its correlation with the exchange bias at the interface. We also point out that we use 2 - 3 orders of magnitude lower in-plane currents compared to references~\onlinecite{Wei2007} and~\onlinecite{Urazhdin2007}, avoiding heating effects and employing a different current path geometry which excludes the possibility of a direct comparison between the experiments.

\section{\label{sec:conclusions}Conclusions}

In conclusion, we have shown that electrical current in the IrMn AFM induces a large torque acting on the adjacent NiFe FM. The torque is in-plane and has an antidamping-like symmetry. We have also shown that there are at least two distinct contributions, one coming from the SHE in IrMn, and the other due to the AFM order of IrMn. The spin-Hall angle of IrMn measured in the sample with the Cu spacer between NiFe and IrMn is found to be $0.056 \pm 0.009 $, comparable to that of Pt. An effective spin-Hall angle of $0.22 \pm 0.04 $, almost three times larger than that of Pt, is measured for the sample with 8~nm IrMn in direct contact with NiFe, exhibiting the largest exchange bias. Our results suggest that electrical current in AFMs can induce torques more efficiently than in most of the heavy NMs. The AFM-induced torques and their correlation with the exchange coupling at the FM/AFM interface could lead to novel designs of spintronic devices. 

\section{Methods and Materials}
\textbf{Materials:} The structures were grown using DC magnetron sputtering on a thermally oxidized Si~(100) substrate. In-plane magnetic field of 200~Oe was applied during growth.

\textbf{Devices:} The microbars are patterned using electron-beam lithography. In Figs.~\ref{fig:resonance-comparison} and \ref{fig:figure5} we show measurements done in bars with 500~nm width and 5~$\mathrm{\mu{m}}$ length, whereas the measurements shown in Fig.~\ref{fig:figure3} are done in bars with 1.8$\times$38~$\mathrm{\mu{m}}$ dimensions. Measurements in Figs.~\ref{fig:resonance-comparison} and \ref{fig:figure3} are repeated in at least two bars with different dimensions. The results are consistent across different bars and all the bar dimensions. The resistivity calibration measurements are done in 4~$\mathrm{\mu{m}}$ wide bars with 40, 80, 120 and 240~$\mathrm{\mu{m}}$ lengths. Typical resistances are on the order of 1~k$\Omega$ for bars with length to width ratios of 10.

\textbf{Experimental procedure:}

For more details on the methods related to our SO-FMR experiments see Refs.~\onlinecite{Fang2011,Kurebayashi2014} and the Supplementary Information therein.

\section{Additional information}
Correspondence and requests for materials should be addressed to AJF (ajf1006@cam.ac.uk).

\begin{acknowledgments}
Authors would like to acknowledge Dr. Hidekazu Kurebayashi and Dr. Tim Skinner for useful discussions. We acknowledge support from the EU European Research Council (ERC) Advanced Grant No. 268066, the Ministry of Education of the Czech Republic Grant No.LM2011026, and the Grant Agency of the Czech Republic Grant No. 14-37427G. VT would like to acknowledge Winton Programme for the Physics of Sustainability and Cambridge Overseas Trusts for financial support.
\end{acknowledgments}

%

\protect\newpage

\onecolumngrid

\makeatletter 
\renewcommand{\thefootnote}{\fnsymbol{footnote}}
\renewcommand{\thefigure}{S\@arabic\c@figure}
\renewcommand{\thesection}{S\@arabic\c@section}
\renewcommand{\theequation}{S.\arabic{equation}}
\renewcommand{\thetable}{S\@arabic\c@table}
\makeatother
\setcounter{section}{0}
\setcounter{equation}{0}
\setcounter{figure}{0}
\setcounter{table}{0}
\setcounter{page}{1}

\begin{center}
	\textbf{\large Supplemental Information: Electrical manipulation of a ferromagnet by an antiferromagnet}\\
	\vspace{0.5cm}
	V. Tshitoyan,$^1$ C. Ciccarelli,$^1$ A. P. Mihai,$^2$ \space M. Ali,$^2$ A. C. Irvine,$^1$ T. A. Moore,$^2$ T. Jungwirth,$^{3,4}$ and A. J. Ferguson$^1$\\
	\vspace{0.5cm}
	$^1$\textit{Microelectronics Group, Cavendish Laboratory, University of Cambridge, CB3 0HE, UK}\\
	$^2$\textit{School of Physics and Astronomy, University of Leeds, Leeds LS2 9JT, UK}\\
	$^3$\textit{Institute of Physics ASCR, v.v.i., Cukrovarnick\'a 10, 162 53 Praha 6, Czech Republic}\\
	$^4$\textit{School of Physics and Astronomy, University of Nottingham, Nottingham NG7 2RD, UK}\\
\end{center}

\section{Microwave current calibration}
\label{sec:microwave-current-calibration}

\begin{figure}[!ht]
	\includegraphics[width=1\columnwidth]{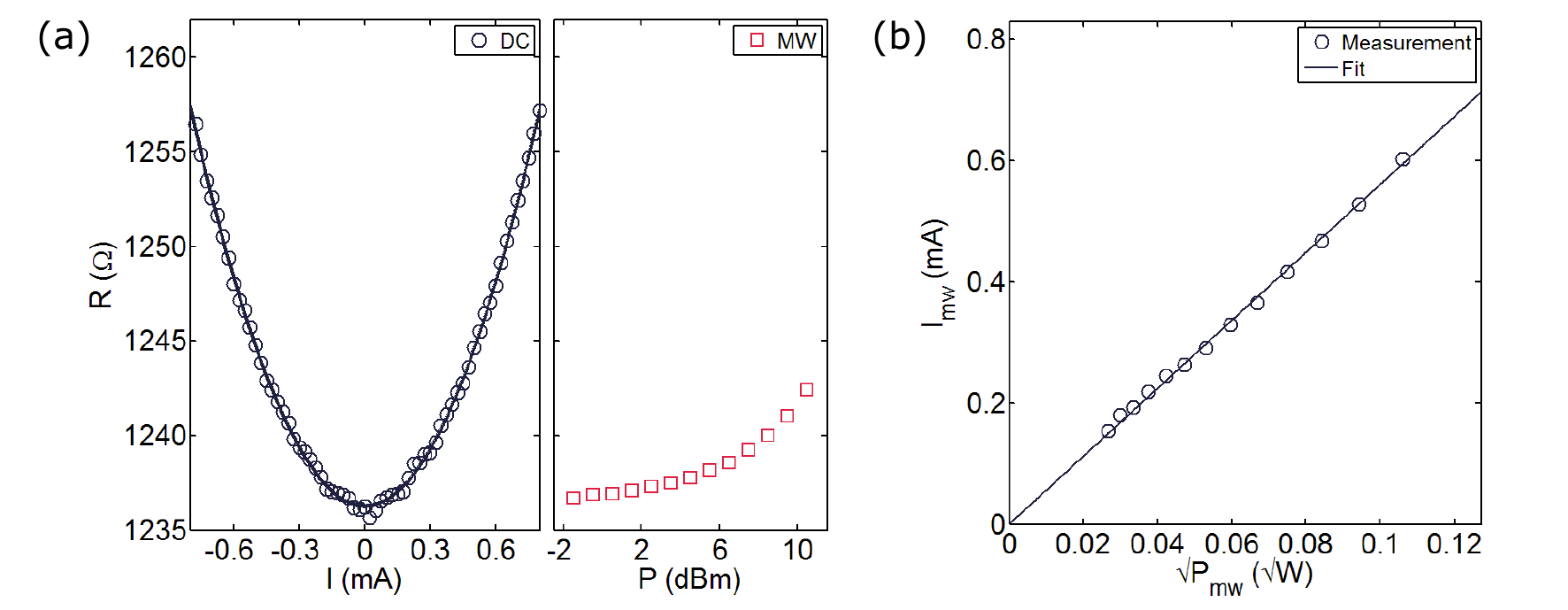}
	\caption{(a) Comparison of resistance change due to heating caused by DC (left) and MW (right) currents. The DC measurement is symmetric with respect to 0 current. (b) MW current vs square root of applied MW power obtained from the heating calibration.}
	\label{fig:heating-calibration}
\end{figure}

Resistances of measured bars vary between a few 100 $\Omega$ and a few $\mathrm{k\Omega}$, thus most of the microwave (MW) power is reflected due to the impedance mismatch between the bar and the MW source ($Z_{out} = 50$ $\Omega$). To calibrate MW current we make use of the Joule heating. The amount of heating is measured using the change of resistance. First DC current is swept from large negative to large positive values and the differential resistance is measured, giving us the resistance change due to DC heating. Then we measure resistance change with increasing microwave power. These measurements for a 500 nm wide and 5 $\mathrm{\mu{m}}$ long bar of $\mathrm{Ru(3)/IrMn(2)/Py(4)/Al(2)}$ are plotted in Fig. \ref{fig:heating-calibration}(a). For DC the value of current is known because it is all dissipated in the bar, there are no reflections. We are able to find the current for each applied MW power by comparing the MW and DC heatings. In Fig. \ref{fig:heating-calibration}(b) we plot values of DC current causing the same amount of heating as MW powers on the x axis. The corresponding MW current is $\sqrt{2}$ times the DC current, because the heating for AC current is given by $I^2R/2$ compared to $I^2R$ for DC (this is already taken into account in the plot). As expected, MW current is linear with the square root of power (in W). From the linear fit we can extract the value of MW current per square root of power.

\section{Layer resistivities}
\label{sec:layer-resistivities}

\begin{figure}[!th]
	\includegraphics[width=1\columnwidth]{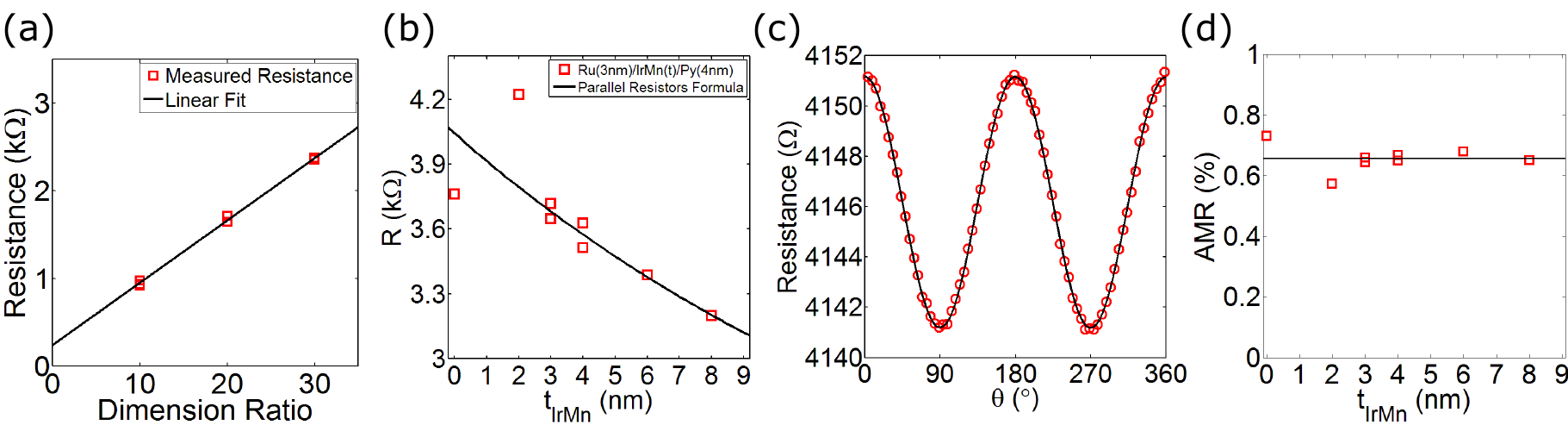}
	\caption{(a) Resistances of bars with different length/width ratios. The fit to a line yields a contact resistance of $235 \pm 75$ $\Omega$. (b) Resistances of bars with different IrMn thicknesses. Resistances of 3 - 8 nm samples are fitted to Eqn. \ref{eqn:parallel-resistors}. (c) Magnetization was rotated in-plane and resistance of the bar was measured. The fit to a $\cos^2\theta$ allows us to determine the AMR magnitude in the multilayer (d) $\mathrm{AMR(\%)} = R_{NiFe}\frac{\Delta{R_{tot}}}{R_{tot}^2}$ for bars with different IrMn thickness, giving us the magnitude of AMR in the NiFe layer.}
	\label{fig:layer-resistivities}
\end{figure}

To know the current distribution in our multilayer stack, which is important for calculations of spin hall angles as well as estimations of the Oersted fields, we deduce resistivities of individual metallic layers. One can not take bulk resistivities because these values change dramatically for thin layers. In addition, there is always an additional contact resistance which has to be taken into account. These values can be determined by a careful analysis of bars with different dimensions and layer thicknesses. In Fig. \ref{fig:layer-resistivities}(a) we plot resistances of 4 $\mathrm{\mu{m}}$ wide bars of 40, 80 and 120 $\mathrm{\mu{m}}$ lengths. The intersection of the linear fit with y axis is the average contact resistance, $R_{cont} = 235 \pm 75$ $\Omega$.

Using the value of contact resistance we can calculate resistivities of individual layers. In Fig. \ref{fig:layer-resistivities}(b) we plot resistances of bars with the same dimensional ratio vs the IrMn thickness t in Ru(3)/IrMn(t)/NiFe(4) structures. The average contact resistance has already been subtracted. Thicknesses are given in nm. We neglect the 2 nm Al capping layer as it is the same for all the samples and is believed to be mainly oxidized. It is surprising that the resistance of the sample with 0 nm IrMn is smaller than that of the sample with 2 nm IrMn. We believe this is due to the higher resistivity of NiFe grown on IrMn compared to that of NiFe grown on Ru. It is know that NiFe can have different resistivities depending on the seed layer ~\cite{Warot2004,Gong2000,Jin2013}. The samples with 3 - 8 nm IrMn fit well to a simple model of parallel resistors, given by 

\begin{equation}
	\label{eqn:parallel-resistors}
	R = R_{cont} + \frac{d\cdot\rho_{IrMn}}{t+d\cdot\rho_{IrMn}/r}.
\end{equation}
Here $d$ is the length/width ratio of the bars (60 for this set of samples), $r$ is the resistance of the multilayer without IrMn, $t$ is the thickness of IrMn and $\rho_{IrMn}$ is its resistivity. The fit in Fig. \ref{fig:layer-resistivities}(b) results in $\rho_{IrMn} = 20.5 \pm 3.5 \times 10^{-7}$ $\Omega \cdot m$.

As already mentioned, resistivity of NiFe is larger if grown on IrMn. This is more prominent for the thinnest (2 nm) IrMn sample. We believe this is due to the worse quality of the 2 nm IrMn interface, as this layer is the thinnest. To calculate different resistivities of NiFe we must know the resistivity of Ru. To estimate the later we use resistances of Ru(3)/IrMn(4)/Ru(2)/NiFe(4) and Ru(3)/NiFe(4), 2624 $\Omega$ and 3760 $\Omega$ respectively. Using the resistivity of IrMn obtained earlier we get $\rho_{Ru} = 4.0 \pm 0.3 \times 10^{-7}$ $\Omega \cdot m$ and $\rho_{NiFe}^{Ru} = 4.7 \pm 0.3 \times 10^{-7}$ $\Omega \cdot m$ for NiFe grown on Ru. Using the resistivity of Ru we can now calculate the resistivity of NiFe grown on IrMn. We use the resistance of Ru(3)/IrMn(2)/NiFe(4) sample and the values obtained from the fit in Fig. \ref{fig:layer-resistivities}(b). We find $\rho_{NiFe}^{2nmIrMn} = 6.9 \pm 0.6 \times 10^{-7}$ $\Omega \cdot m$ and $\rho_{NiFe}^{IrMn} = 5.4 \pm 0.4 \times 10^{-7}$ $\Omega \cdot m$.

To verify the parallel resistors approach, we compare values of AMR for layers with different IrMn thicknesses. Change of the resistance due to AMR is extracted by rotating the direction of the magnetic field with respect to the sample. A typical measurement result is plotted in Fig. \ref{fig:layer-resistivities}(c). The value of measured AMR depends on the proportion of the current in the NiFe layer, and the size of its AMR. For example for thicker IrMn samples the proportion of the current in NiFe is smaller and thus smaller total AMR is measured. We deduce the exact relationship using the parallel resistors model.

\begin{equation}
	R_{tot} = (1/R_{NiFe} + 1/R_{rest})^{-1} = \frac{R_{NiFe}R_{rest}}{R_{NiFe} + R_{rest}}
\end{equation}
\begin{equation}
	\begin{split}
		\Delta{R_{tot}} & =  \frac{(R_{NiFe} + \Delta{R_{NiFe}})R_{rest}}{R_{NiFe} + \Delta{R_{NiFe}} + R_{rest}} - \frac{R_{NiFe}R_{rest}}{R_{NiFe} + R_{rest}} = \\
		& = \frac{\Delta{R_{NiFe}}R_{rest}^2}{(R_{NiFe} + \Delta{R_{NiFe}} + R_{rest})(R_{NiFe} + R_{rest})}
	\end{split}
\end{equation}
\begin{equation}
	\frac{\Delta{R_{tot}}}{R_{tot}} = \frac{\Delta{R_{NiFe}}R_{rest}^2}{(R_{NiFe} + \Delta{R_{NiFe}} + R_{rest})(R_{NiFe} + R_{rest})} \cdot \frac{R_{NiFe} + R_{rest}}{R_{NiFe}R_{rest}} \approx \Delta{R_{NiFe}}\frac{R_{tot}}{R_{NiFe}^2}
\end{equation}
\begin{equation}
	\frac{\Delta{R_{tot}}}{R_{tot}^2} \approx \frac{\Delta{R_{NiFe}}}{R_{NiFe}^2}
\end{equation}

Thus $\frac{\Delta{R_{NiFe}}}{R_{NiFe}} = R_{NiFe}\frac{\Delta{R_{tot}}}{R_{tot}^2}$ gives us the value of AMR in NiFe. In Fig. \ref{fig:layer-resistivities}(d) we plot the right hand side of this equation for all measured IrMn thicknesses. As one can see it is almost the same for the 3 - 8 nm thickness range of IrMn, and is approximately 0.7 \%. For the sample with 2 nm IrMn AMR of NiFe is slightly smaller, whereas it is slightly larger for NiFe grown on Ru, as expected due to worse and better layer qualities respectively. Decrease of intrinsic AMR of NiFe for thin layers, as well as its dependence on the seed layer has been reported previously ~\cite{Yeh1987,Rijks1995,Gong2000,Choe1999}. 

We believe that the agreement of AMR magnitudes, in combination with the relatively good fit of IrMn thickness dependence of sample resistances, implies that our parallel resistors approach is valid and estimates of layer resistivities are correct. As yet another additional supporting argument for our calculation, the bulk resistivity ratio is approximately 18(IrMn):1(Ru):2(NiFe).($1260 \times 10^{-7}$ $\Omega \cdot m$, $71 \times 10^{-7}$ $\Omega \cdot m$ , $140 \times 10^{-7}$ $\Omega \cdot m$) ~\cite{Acharyya2012,Choi2002,Mayadas1974}. The ratios of resistivities deduced above are 5.1 : 1 : 1.2-1.8. The order is the same, but differences in resistivities are more moderate for thin films because scattering off interfaces is substantial, and thus resistivity must be less material-dependent.

\begin{figure}[!ht]
	\centering
	\includegraphics[width=0.45\textwidth]{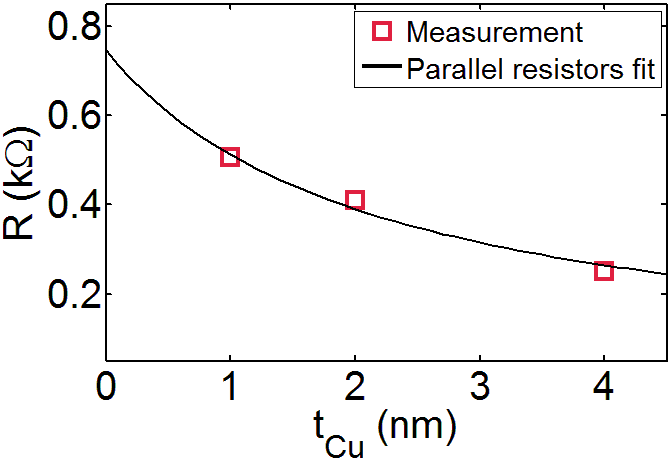}
	\caption{Resistances of bars with different Cu spacer thicknesses fitted to the parallel resistors formula.}
	\label{fig:spin-hall-angles-1}
\end{figure}

The resistivity of Cu is deduced from the Ru(3)/IrMn(4)/Cu($\mathrm{t_{Cu}}$)/Py(4) structures , where $\mathrm{t_{Cu}}$ is 1, 2 or 4 nm. In Fig. \ref{fig:spin-hall-angles-1} we plot resistances of bars with different Cu thicknesses fitted to equation \ref{eqn:parallel-resistors}, except instead of $\rho_{IrMn}$ one has $\rho_{Cu}$. From the fit we find $\rho_{Cu} = 1.55\cdot10^{-7}$ $\Omega{m}$.

\section{Oersted field}
\label{sec:oersted-field}

\begin{figure}[!t]
	\centering
	\includegraphics[width=0.65\textwidth]{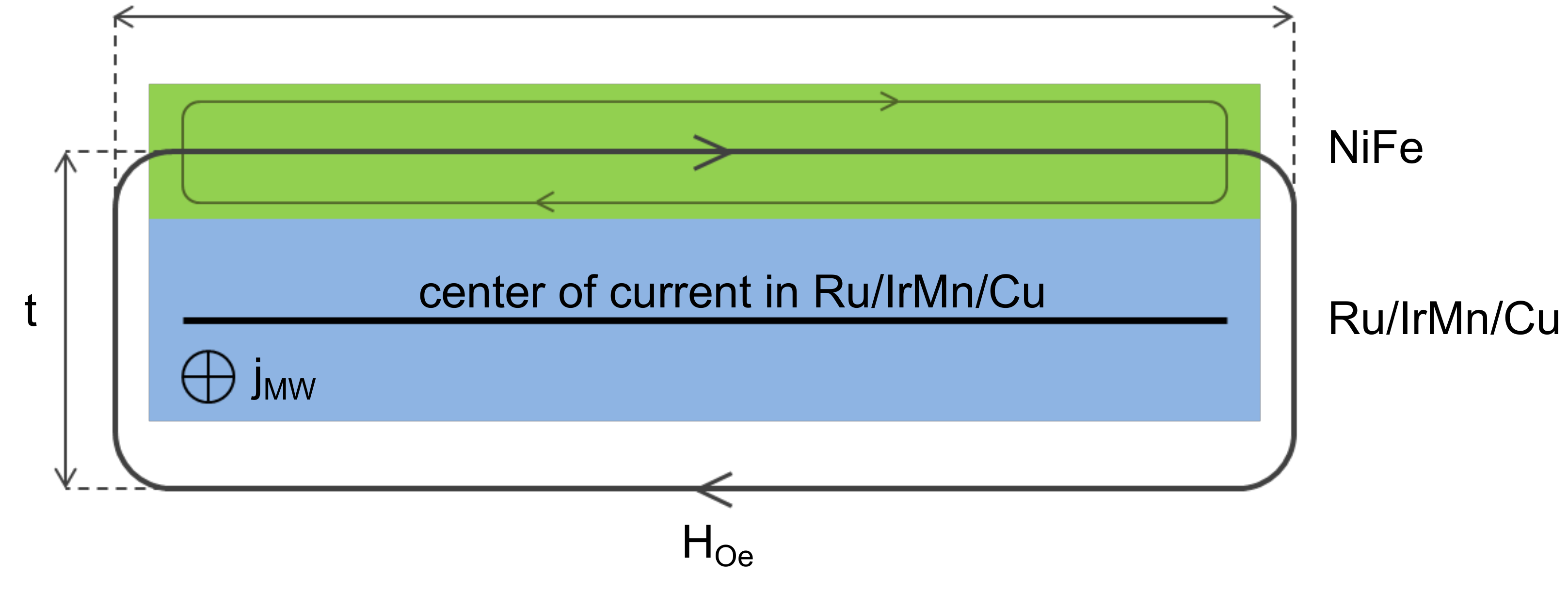}
	\caption{Schematic representation of Oersted fields induced by the current in the multilayer.}
	\label{fig:oersted-field}
\end{figure}

Current in the IrMn, Ru and Cu layers creates an effective Oersted field in y direction at the centre of the NiFe layer. Current in the NiFe itself generates only a symmetric Oersted field with respect to the centre of the layer which does not contribute to the effective $h_{y}$ or $h_{z}$ (Fig. \ref{fig:oersted-field}). From Ampere's law we have

\begin{equation}
	\oint{H_{Oe}\mathrm{d}l} = I
\end{equation}
Where $I$ is the current encircled by the integration loop. For our geometry sketched in Fig. \ref{fig:oersted-field}(a) we can write

\begin{equation}
	\mu_{0}H_{Oe} = \frac{\mu_{0}I_{Oe}}{2(w+t)} \approx \frac{\mu_{0}I_{Oe}}{2w}.
	\label{eqn:oersted-field}
\end{equation}
Here $I_{Oe}$ is the current in the Ru, IrMn and Cu layers. We used the fact that the thickness $t$ of the bar ($\sim$ 10 nm) is very small compared to its width $w$ (500 nm - 4 $\mu{m}$) for all measured devices. This means that the Oersted field depends only on the size of the current in Ru, IrMn and Cu layers, and not on layer thicknesses, similar to the case of an infinite plane.

\section{Magnetic anisotropies: $A_{sym}$ and $A_{asy}$}
\label{sec:magnetic-anisotropies}

Total magnetic anisotropy is modelled as a combination of unidirectional, uniaxial and rotational anisotropies. Unidirectional anisotropy models the exchange bias. Uniaxial anisotropy is a combination of shape anisotropy, crystalline anisotropy of NiFe and some uniaxial anisotropy due to the exchange bias ~\cite{Stiles1999}. The contribution of each of these towards the cumulative uniaxial anisotropy varies depending on the dimensions of the bar and the thickness of the IrMn layer. Rotational anisotropy is due to the partially stable grains of the polycristalline IrMn coupling to the NiFe at the interface. These are the same AFM grains responsible for the increased coercivity of magnetic hysteresis measurements ~\cite{Stiles1999}. This anisotropy is modelled as an additional in-plane effective field $H_{rot}$ along the NiFe magnetizatoin direction. Magnetic free energy per unit area becomes

\begin{equation}
	\begin{split}
		F[\theta, \phi] & = F_{Zeeman}[\theta,\phi] + F_{surf}[\theta,\phi] + F_{shape}[\theta,\phi] + F_{U}[\theta,\phi] + F_{exch}[\theta,\phi] = \\
		&-\mu_0(H+H_{rot})Mt_{FM}(\sin{\phi} \sin{\phi_H}\cos(\theta-\theta_H)+\cos{\phi} \cos{\phi_H}) \\ 
		&+ (\mu_0{M}^2t_{FM}/2-K_{S})\cos^2\phi-K_Ut_{FM}\sin^2\phi{\cos^2(\theta-\theta_{uni})} \\ 
		&- \mu_0Mt_{FM}H_{ex}\cos(\theta - \theta_{exch})\sin\phi,
	\end{split}
\end{equation}
where ($\theta_H$, $\phi_H$) and ($\theta$, $\phi$) are in and out-of-plane angles of applied field $H$ and magnetization $M$ in spherical coordinates, with $\phi = 90^\circ$ being in the plane of the sample. $K_S$ and $K_U$ are surface and in-plane uniaxial anisotropy constants, $t_{FM}$ is the thickness of the ferromagnetic layer, $H_{ex}$ is the exchange bias field, $\theta_{uni}$ and $\theta_{exch}$ are directions of the uniaxial anisotropy and the exchange bias respectively. The resonance condition reads

\begin{equation}
	\left(\frac{\omega}{\gamma}\right)^2=\frac{1}{M^2t_{FM}^2\sin^2\theta}\cdot\left[\left(\frac{\partial^2F}{\partial\theta^2}\right)\left(\frac{\partial^2F}{\partial\phi^2}\right) - \left(\frac{\partial^2F}{\partial\phi\partial\theta}\right)^2\right],
\end{equation}
where $\omega$ is the resonance frequency and $\gamma$ is the gyromagnetic ratio. Plugging in the expression for $F[\theta, \phi]$ into the equation above and differentiating it with respect to $\theta$ and $\phi$ one obtains

\begin{equation}
	\label{eqn:kittel-modified}
	\left(\frac{\omega}{\gamma}\right)^2 = \mu_0^2(H + H_1)(H + H_2)
\end{equation}
with

\begin{equation}
	\label{eqn:h1-and-h2}
	\begin{split}
		H_1 & = H_{rot} + M_{eff} + H_{exch}\cos(\theta - \theta_{exch}) + H_U\cos^2(\theta - \theta_{U}) \\
		H_2 & = H_{rot} + H_{exch}\cos(\theta - \theta{exch}) + H_U\cos[2(\theta - \theta_{U})].
	\end{split}
\end{equation}
Here we have relabelled variables in the following way

\begin{equation}
	\label{eqn:meff}
	\begin{split}
		M_{eff} & = M - 2K_S/\mu_0Mt_{FM} \\
		H_U & = 2K_U/\mu_0M.
	\end{split}
\end{equation}

We used these equations to fit the $\theta$ dependence of the resonance field and extract anisotropies of each sample. In this model $M_{eff}$ and $H_{rot}$ are correlated, thus we need to know one of these using a different method. This correlation is easier to see if we rewrite equation \ref{eqn:kittel-modified} making an approximation $H_{res} + H_1 \approx M_{eff}$. This is valid because the rest of the terms in $H_1$ are much smaller than $M_{eff}$. We write \ref{eqn:kittel-modified} as

\begin{equation}
	\mu_0H_{res} = \left(\frac{\omega}{\gamma}\right)^2 \frac{1}{\mu_0M_{eff}} - \mu_0H_{rot} - \mu_0H_{exch}\cos(\theta - \theta_{exch}) - \mu_0H_U\cos[2(\theta - \theta{U})].
\end{equation}

For the given frequency larger $M_{eff}$ leads to a smaller $H_{rot}$ and vice versa, so we extract $M_{eff}$ from the frequency dependence of the resonance field and use it to fit out $H_{rot}$ (the fitting is done using the full model and not the approximation).

$A_{sym}$ and $A_{asy}$ entering the expressions for the rectified dc voltage are given by

\begin{equation}
	\label{eqn:asym-and-aasy}
	\begin{split}
		A_{sym} & = \frac{\gamma(H_{res}+H_1)(H_{res}+H_2)}{\omega\Delta{H}(2H_{res} + H_1 + H_2)} \\
		A_{asy} & = \frac{(H_{res}+H_1)}{\mu_0\Delta{H}(2H_{res} + H_1 + H_2)}
	\end{split}
\end{equation}

as deduced in reference ~\cite{Fang2011}, with $H_1$ and $H_2$ given by equations \ref{eqn:h1-and-h2}, and $\Delta{H}$ being the resonance linewidth.

\begin{figure}[!b]
	\centering
	\includegraphics[height=0.4\textwidth]{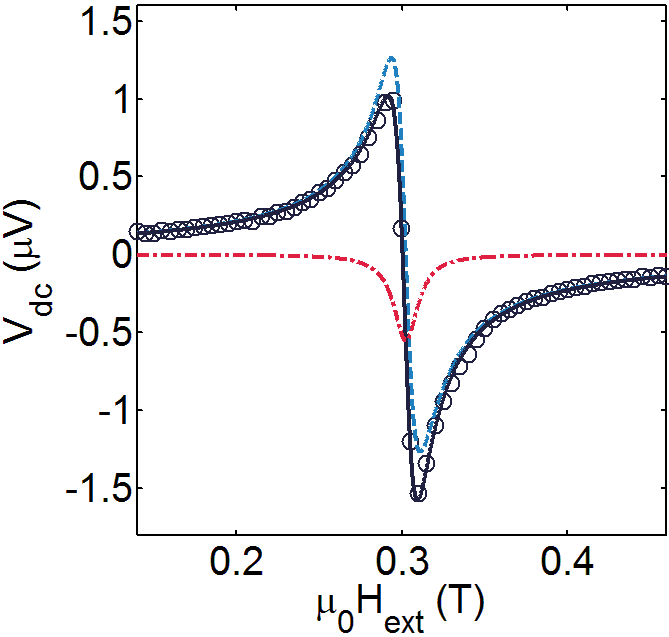}
	\caption{(a) A resonance curve measured in the Ru(3)Py(4) bar at 17.9 GHz, decomposed into symmetric and antisymmetric Lorentzians.}
	\label{fig:spin-hall-angle-of-ru}
\end{figure}

\section{Spin-Hall angle of Ru}
\label{sec:spin-hall-angle-of-ru}

The spin hall angle of Ru is calculated using the $h_{z}/h_{y}$ ratio measured experimentally in a bar patterned from the Ru(3)Py(4) bilayer ~\cite{Liu2011}, assuming that $h_y$ is predominantly due to the Oersted field. We use 

\begin{equation}
	\label{eqn:spin-hall-angle-ru}
	\theta_{SH} = \frac{h_z}{h_y}\cdot\frac{e\mu_0M_st_{Ru}t_{NiFe}}{\hbar}.
\end{equation}
Here e is the electron charge, $\mu_0M_s = 1$ T for NiFe, $t_{Ru} = 3$ nm, $t_{NiFe} = 4$ nm, $\hbar$ is the reduced Planck constant. A typical resonance measurement in this sample is shown in Fig. \ref{fig:spin-hall-angle-of-ru}. We find

\begin{equation}
	\theta_{SH} =(0.50 \pm 0.02)\cdot\frac{1.602{\cdot}10^{-19}{\cdot}1\cdot3\cdot4\cdot10^{-18}}{1.055\cdot10^{-34}} = 0.0092 \pm 0.0004.
\end{equation}

\section{Samples with Ta seed layers}
\label{sec:samples-with-a-ta-seed-layer}

\begin{figure}[!h]
	\includegraphics[width=1\columnwidth]{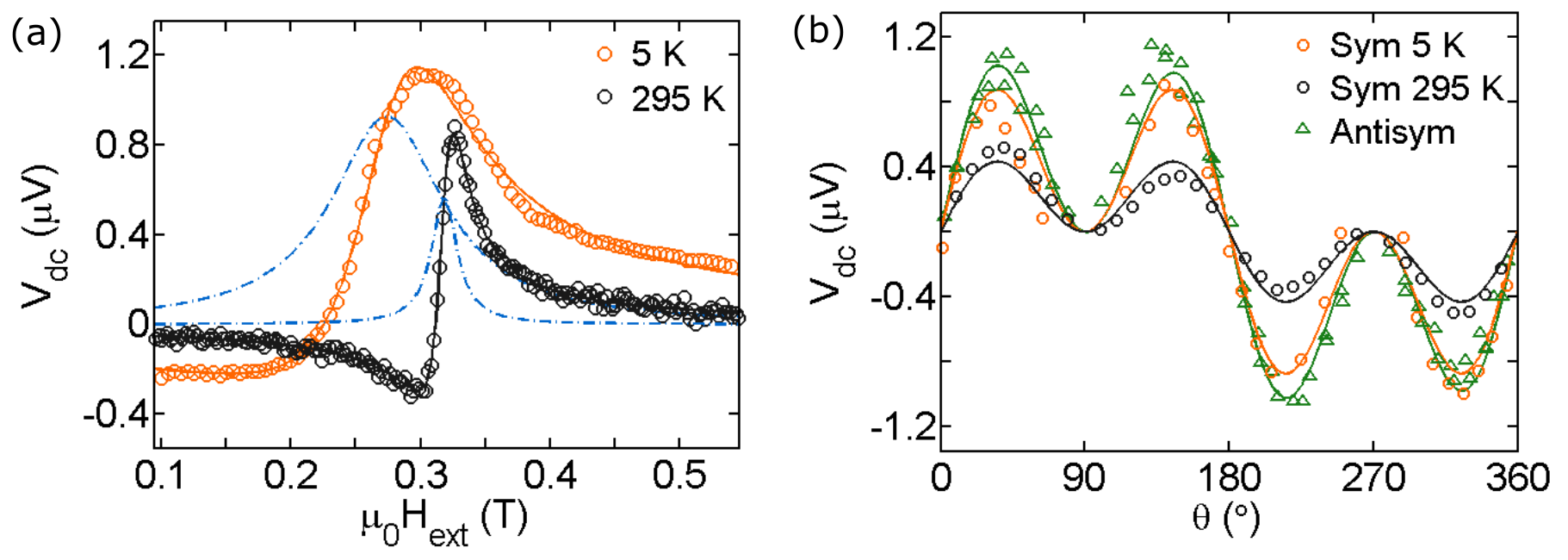}
	\caption{(a) Resonances measured for the 3 nm IrMn sample at 5 K and for the 2 nm IrMn sample at room temperature (295K), at 16.5 and 18.6 GHz microwave frequencies respectively. The antisymmetric components are normalized to 1~$\mathrm{\mu{V}}$ (not shown), and the symmetric components are show with dotted lines. (b) Angle dependences of the symmetric and the antisymmetric components for the measurements shown in (a). Solid lines are fits to $h_z\sin2\theta\cos\theta$ (symmetric) and $\sin2\theta(h_y\cos\theta - h_x\sin\theta$) (antisymmetric).}
	\label{fig:ta-resonances}
\end{figure}

To confirm the fact that the seed layer does not have a major contribution to the observed anti-damping torque we measure $\mathrm{SiO_{x}/Ta(4.5)/IrMn(2,3)/NiFe(4)/Al(2)}$ structures with a 4.5~nm Ta seed layer instead of Ru, both at room temperature and at 5 K. Neither 2 nor 3 nm IrMn samples exhibit exchange bias at room temperature. The 2 nm IrMn sample does not develop any substantial exchange bias even at low temperatures, whereas the 3 nm IrMn sample develops an exchange bias of 8~$\pm$~1~mT at 5~K. In Fig.~\ref{fig:ta-resonances}(a) we plot resonances measured for the 2 nm IrMn sample at room temperature and for the 3 nm IrMn sample at 5 K. Firstly, in both cases the symmetric component is positive. Ta has a large negative spin-Hall angle and if the effect was dominated by the spin-current from Ta one would expect $h_z$ and thus the symmetric component to be negative for a positive antisymmetric component. The fact that $h_z$ is positive means that any effects due to the spin-Hall effect in Ta are small compared to the IrMn-induced effects. Additionally, one can see that at low temperature the symmetric component becomes even larger. This can also be clearly seen in the angle dependences of $V_{sym}$ and $V_{asy}$ plotted in Fig.~\ref{fig:ta-resonances}(b). This result further supports the argument that the increase of the anti-damping torque with the exchange bias observed in our experiments is not related to the efficiency of the transfer of the spin-angular momentum generated in the seed layer, as this would lead to a decrease of $h_z$ for a seed layer with a negative spin-Hall angle like Ta. We believe that in our experiments the spin angular momentum generated in the seed layer is fully absorbed by the first few atomic layers of IrMn due to its small spin diffusion length, and the observed anti-damping torque is induced predominantly by the antiferromagnet.

\newpage

\section{Independence of the symmetry of $h_z$ on the exchange bias direction}
\label{sec:symmetries}

\begin{figure}[h]
	\centering
	\includegraphics[width=0.95\textwidth]{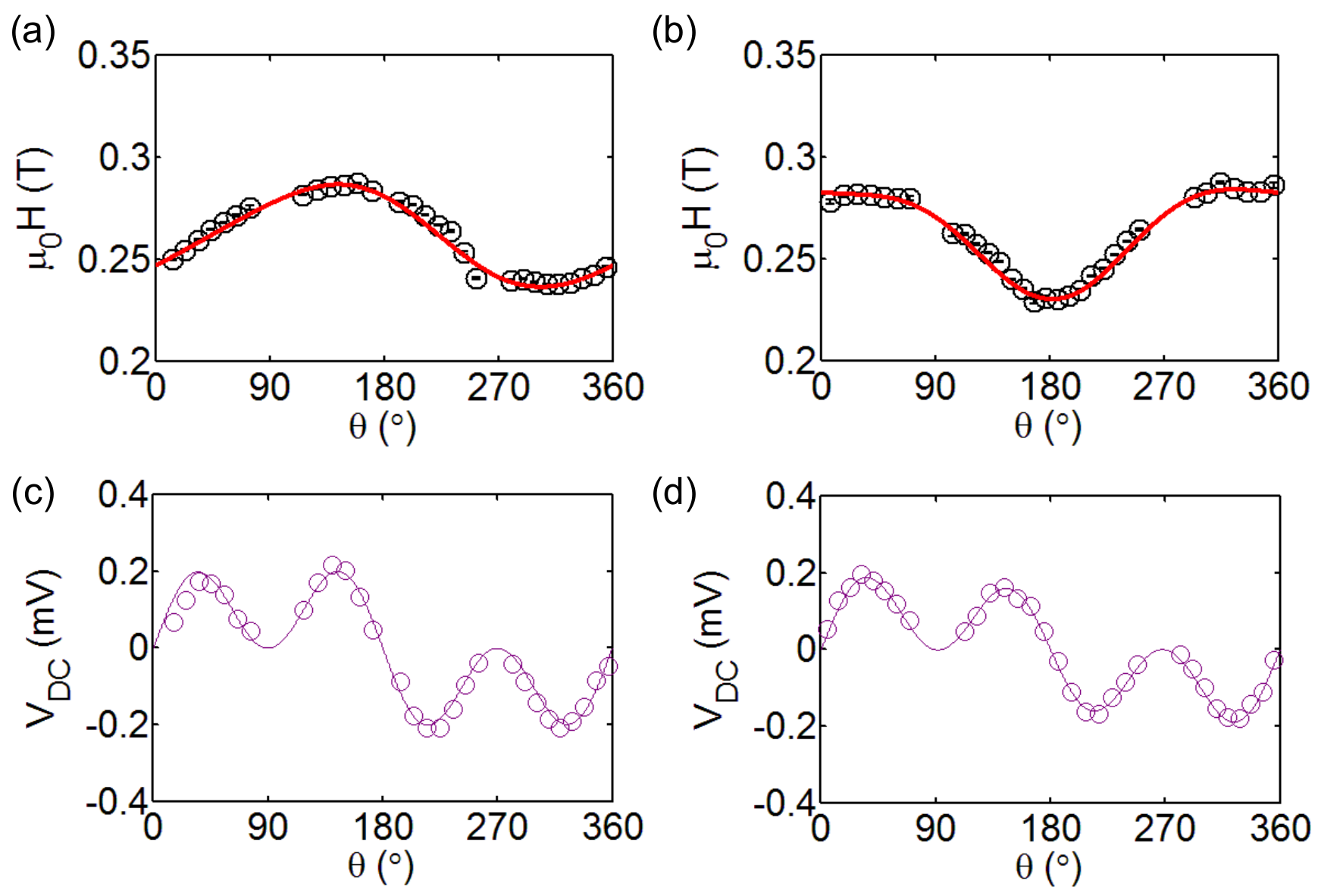}
	\caption{(a, b) Angle dependence of the resonance field for two different bars with 4 nm IrMn and (c, d) corresponding symmetric components of the measured dc voltage. Although the exchange bias is substantial and has different directions for the two bars, the angle dependence of the symmetric component of the Lorentzian, which corresponds to $h_z$, is not affected. }
	\label{fig:symmetries}
\end{figure}

\section{Power, frequency and dimension dependence of current-induced torques}
\label{sec:power-frequency-geometry-dependence}

\begin{figure}[!h]
	\centering
	\includegraphics[width=0.98\textwidth]{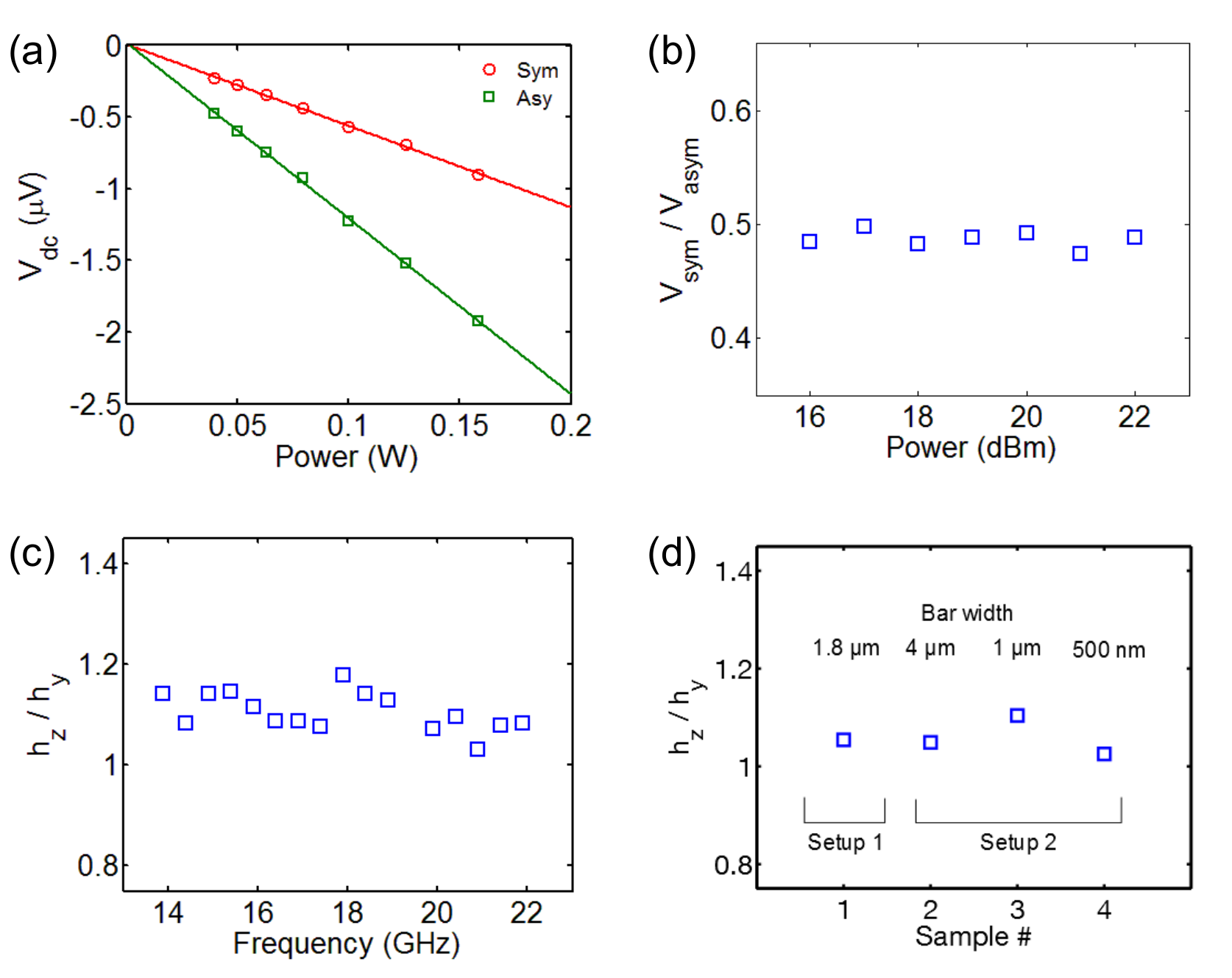}
	\caption{(a) Dependence of the magnitudes of symmetric and antisymmetric Lorentzians on applied microwave power for the 2 nm IrMn sample, fitted to lines. (b) Microwave power dependence of the ratio of symmetric and antisymmetric Lorentzians plotted in (a). (c) Frequency dependence of the $h_z/h_y$ ratio for the 3 nm IrMn sample. (d) $h_z/h_y$ ratio for 2 nm IrMn samples with different dimensions and measured in two different setups. The bar dimensions are 1.8 $\mathrm{\mu{m}}$ x 38 $\mathrm{\mu{m}}$, 4 $\mathrm{\mu{m}}$ x 240 $\mathrm{\mu{m}}$, 1 $\mathrm{\mu{m}}$ x 10 $\mathrm{\mu{m}}$, 500 nm x 5 $\mathrm{\mu{m}}$}
	\label{fig:freq-and-power-dep}
\end{figure}

We present several control measurements to support our interpretation of symmetric and antisymmetric components of the resonance. In Fig.~\ref{fig:freq-and-power-dep}(a) we show the power dependence of the FMR magnitude for the 2 nm IrMn sample at room temperature, measured at 17.9 GHz. As one can see, both symmetric and antisymmetric components scale linearly with power, as expected for the rectification signal (h, I $\propto$ $\sqrt{P}$, see equations 1 and 2 in the main text). Their ratio is power independent (Fig.~\ref{fig:freq-and-power-dep}(b)). 

In Fig.~\ref{fig:freq-and-power-dep}(c) we show that the $h_z$ / $h_y$ ratio is frequency independent in our devices. The data shown is for the 3 nm IrMn sample. Note that here the ratio is extracted from single resonances rather than a full angle-dependent measurement, thus the relatively large fluctuations, although still within about 10 \% of each other.

The measurements were performed in bars with different dimensions to exclude any geometry related effects. Parts of measurements were also performed in two different measurement systems, with the same results. Fig.~\ref{fig:freq-and-power-dep}(d) summarizes the above stated for the 2 nm IrMn sample.

\begin{thebibliography}{43}%
	\makeatletter
	\providecommand \@ifxundefined [1]{%
		\@ifx{#1\undefined}
	}%
	\providecommand \@ifnum [1]{%
		\ifnum #1\expandafter \@firstoftwo
		\else \expandafter \@secondoftwo
		\fi
	}%
	\providecommand \@ifx [1]{%
		\ifx #1\expandafter \@firstoftwo
		\else \expandafter \@secondoftwo
		\fi
	}%
	\providecommand \natexlab [1]{#1}%
	\providecommand \enquote  [1]{``#1''}%
	\providecommand \bibnamefont  [1]{#1}%
	\providecommand \bibfnamefont [1]{#1}%
	\providecommand \citenamefont [1]{#1}%
	\providecommand \href@noop [0]{\@secondoftwo}%
	\providecommand \href [0]{\begingroup \@sanitize@url \@href}%
	\providecommand \@href[1]{\@@startlink{#1}\@@href}%
	\providecommand \@@href[1]{\endgroup#1\@@endlink}%
	\providecommand \@sanitize@url [0]{\catcode `\\12\catcode `\$12\catcode
		`\&12\catcode `\#12\catcode `\^12\catcode `\_12\catcode `\%12\relax}%
	\providecommand \@@startlink[1]{}%
	\providecommand \@@endlink[0]{}%
	\providecommand \url  [0]{\begingroup\@sanitize@url \@url }%
	\providecommand \@url [1]{\endgroup\@href {#1}{\urlprefix }}%
	\providecommand \urlprefix  [0]{URL }%
	\providecommand \Eprint [0]{\href }%
	\providecommand \doibase [0]{http://dx.doi.org/}%
	\providecommand \selectlanguage [0]{\@gobble}%
	\providecommand \bibinfo  [0]{\@secondoftwo}%
	\providecommand \bibfield  [0]{\@secondoftwo}%
	\providecommand \translation [1]{[#1]}%
	\providecommand \BibitemOpen [0]{}%
	\providecommand \bibitemStop [0]{}%
	\providecommand \bibitemNoStop [0]{.\EOS\space}%
	\providecommand \EOS [0]{\spacefactor3000\relax}%
	\providecommand \BibitemShut  [1]{\csname bibitem#1\endcsname}%
	\let\auto@bib@innerbib\@empty
	\bibitem [{\citenamefont {N\'{u}\~{n}ez}\ \emph {et~al.}(2006)\citenamefont
		{N\'{u}\~{n}ez}, \citenamefont {Duine}, \citenamefont {Haney},\ and\
		\citenamefont {MacDonald}}]{Nunez2006}%
	\BibitemOpen
	\bibfield  {author} {\bibinfo {author} {\bibfnamefont {A.}~\bibnamefont
			{N\'{u}\~{n}ez}}, \bibinfo {author} {\bibfnamefont {R.}~\bibnamefont
			{Duine}}, \bibinfo {author} {\bibfnamefont {P.}~\bibnamefont {Haney}}, \ and\
		\bibinfo {author} {\bibfnamefont {A.}~\bibnamefont {MacDonald}},\ }\href
	{\doibase 10.1103/PhysRevB.73.214426} {\bibfield  {journal} {\bibinfo
			{journal} {Physical Review B}\ }\textbf {\bibinfo {volume} {73}},\ \bibinfo
		{pages} {1} (\bibinfo {year} {2006})}\BibitemShut {NoStop}%
	\bibitem [{\citenamefont {Haney}\ and\ \citenamefont
		{MacDonald}(2008)}]{Haney2008}%
	\BibitemOpen
	\bibfield  {author} {\bibinfo {author} {\bibfnamefont {P.}~\bibnamefont
			{Haney}}\ and\ \bibinfo {author} {\bibfnamefont {A.}~\bibnamefont
			{MacDonald}},\ }\href {\doibase 10.1103/PhysRevLett.100.196801} {\bibfield
		{journal} {\bibinfo  {journal} {Physical Review Letters}\ }\textbf {\bibinfo
			{volume} {100}},\ \bibinfo {pages} {1} (\bibinfo {year} {2008})}\BibitemShut
	{NoStop}%
	\bibitem [{\citenamefont {Xu}\ \emph {et~al.}(2008)\citenamefont {Xu},
		\citenamefont {Wang},\ and\ \citenamefont {Xia}}]{Xu2008}%
	\BibitemOpen
	\bibfield  {author} {\bibinfo {author} {\bibfnamefont {Y.}~\bibnamefont
			{Xu}}, \bibinfo {author} {\bibfnamefont {S.}~\bibnamefont {Wang}}, \ and\
		\bibinfo {author} {\bibfnamefont {K.}~\bibnamefont {Xia}},\ }\href {\doibase
		10.1103/PhysRevLett.100.226602} {\bibfield  {journal} {\bibinfo  {journal}
			{Physical Review Letters}\ }\textbf {\bibinfo {volume} {100}},\ \bibinfo
		{pages} {1} (\bibinfo {year} {2008})}\BibitemShut {NoStop}%
	\bibitem [{\citenamefont {Gomonay}(2010)}]{Gomonay2010}%
	\BibitemOpen
	\bibfield  {author} {\bibinfo {author} {\bibfnamefont {H.~V.}\ \bibnamefont
			{Gomonay}},\ }\href {\doibase 10.1103/PhysRevB.81.144427} {\bibfield
		{journal} {\bibinfo  {journal} {Physical Review B}\ }\textbf {\bibinfo
			{volume} {81}},\ \bibinfo {pages} {1} (\bibinfo {year} {2010})}\BibitemShut
	{NoStop}%
	\bibitem [{\citenamefont {Hals}\ \emph {et~al.}(2011)\citenamefont {Hals},
		\citenamefont {Tserkovnyak},\ and\ \citenamefont {Brataas}}]{Hals2011}%
	\BibitemOpen
	\bibfield  {author} {\bibinfo {author} {\bibfnamefont {K.}~\bibnamefont
			{Hals}}, \bibinfo {author} {\bibfnamefont {Y.}~\bibnamefont {Tserkovnyak}}, \
		and\ \bibinfo {author} {\bibfnamefont {A.}~\bibnamefont {Brataas}},\ }\href
	{\doibase 10.1103/PhysRevLett.106.107206} {\bibfield  {journal} {\bibinfo
			{journal} {Physical Review Letters}\ }\textbf {\bibinfo {volume} {106}},\
		\bibinfo {pages} {1} (\bibinfo {year} {2011})}\BibitemShut {NoStop}%
	\bibitem [{\citenamefont {Shick}\ \emph {et~al.}(2010)\citenamefont {Shick},
		\citenamefont {Khmelevskyi}, \citenamefont {Mryasov}, \citenamefont
		{Wunderlich},\ and\ \citenamefont {Jungwirth}}]{Shick2010}%
	\BibitemOpen
	\bibfield  {author} {\bibinfo {author} {\bibfnamefont {A.~B.}\ \bibnamefont
			{Shick}}, \bibinfo {author} {\bibfnamefont {S.}~\bibnamefont {Khmelevskyi}},
		\bibinfo {author} {\bibfnamefont {O.~N.}\ \bibnamefont {Mryasov}}, \bibinfo
		{author} {\bibfnamefont {J.}~\bibnamefont {Wunderlich}}, \ and\ \bibinfo
		{author} {\bibfnamefont {T.}~\bibnamefont {Jungwirth}},\ }\href {\doibase
		10.1103/PhysRevB.81.212409} {\bibfield  {journal} {\bibinfo  {journal}
			{Physical Review B}\ }\textbf {\bibinfo {volume} {81}},\ \bibinfo {pages}
		{212409} (\bibinfo {year} {2010})}\BibitemShut {NoStop}%
	\bibitem [{\citenamefont {\v{Z}elezn\'{y}}\ \emph {et~al.}(2014)\citenamefont
		{\v{Z}elezn\'{y}}, \citenamefont {Gao}, \citenamefont {V\'{y}born\'{y}},
		\citenamefont {Zemen}, \citenamefont {Ma\v{s}ek}, \citenamefont {Manchon},
		\citenamefont {Wunderlich}, \citenamefont {Sinova},\ and\ \citenamefont
		{Jungwirth}}]{Zelezny2014}%
	\BibitemOpen
	\bibfield  {author} {\bibinfo {author} {\bibfnamefont {J.}~\bibnamefont
			{\v{Z}elezn\'{y}}}, \bibinfo {author} {\bibfnamefont {H.}~\bibnamefont
			{Gao}}, \bibinfo {author} {\bibfnamefont {K.}~\bibnamefont
			{V\'{y}born\'{y}}}, \bibinfo {author} {\bibfnamefont {J.}~\bibnamefont
			{Zemen}}, \bibinfo {author} {\bibfnamefont {J.}~\bibnamefont {Ma\v{s}ek}},
		\bibinfo {author} {\bibfnamefont {A.}~\bibnamefont {Manchon}}, \bibinfo
		{author} {\bibfnamefont {J.}~\bibnamefont {Wunderlich}}, \bibinfo {author}
		{\bibfnamefont {J.}~\bibnamefont {Sinova}}, \ and\ \bibinfo {author}
		{\bibfnamefont {T.}~\bibnamefont {Jungwirth}},\ }\href {\doibase
		10.1103/PhysRevLett.113.157201} {\bibfield  {journal} {\bibinfo  {journal}
			{Physical Review Letters}\ }\textbf {\bibinfo {volume} {113}},\ \bibinfo
		{pages} {157201} (\bibinfo {year} {2014})}\BibitemShut {NoStop}%
	\bibitem [{\citenamefont {Nogu\'{e}s}\ and\ \citenamefont
		{Schuller}(1999)}]{Nogues1999}%
	\BibitemOpen
	\bibfield  {author} {\bibinfo {author} {\bibfnamefont {J.}~\bibnamefont
			{Nogu\'{e}s}}\ and\ \bibinfo {author} {\bibfnamefont {I.~K.}\ \bibnamefont
			{Schuller}},\ }\href {\doibase 10.1016/S0304-8853(98)00266-2} {\bibfield
		{journal} {\bibinfo  {journal} {Journal of Magnetism and Magnetic Materials}\
		}\textbf {\bibinfo {volume} {192}},\ \bibinfo {pages} {203} (\bibinfo {year}
		{1999})}\BibitemShut {NoStop}%
	\bibitem [{\citenamefont {Park}\ \emph {et~al.}(2011)\citenamefont {Park},
		\citenamefont {Wunderlich}, \citenamefont {Mart\'{\i}}, \citenamefont
		{Hol\'{y}}, \citenamefont {Kurosaki}, \citenamefont {Yamada}, \citenamefont
		{Yamamoto}, \citenamefont {Nishide}, \citenamefont {Hayakawa}, \citenamefont
		{Takahashi}, \citenamefont {Shick},\ and\ \citenamefont
		{Jungwirth}}]{Park2011b}%
	\BibitemOpen
	\bibfield  {author} {\bibinfo {author} {\bibfnamefont {B.~G.}\ \bibnamefont
			{Park}}, \bibinfo {author} {\bibfnamefont {J.}~\bibnamefont {Wunderlich}},
		\bibinfo {author} {\bibfnamefont {X.}~\bibnamefont {Mart\'{\i}}}, \bibinfo
		{author} {\bibfnamefont {V.}~\bibnamefont {Hol\'{y}}}, \bibinfo {author}
		{\bibfnamefont {Y.}~\bibnamefont {Kurosaki}}, \bibinfo {author}
		{\bibfnamefont {M.}~\bibnamefont {Yamada}}, \bibinfo {author} {\bibfnamefont
			{H.}~\bibnamefont {Yamamoto}}, \bibinfo {author} {\bibfnamefont
			{A.}~\bibnamefont {Nishide}}, \bibinfo {author} {\bibfnamefont
			{J.}~\bibnamefont {Hayakawa}}, \bibinfo {author} {\bibfnamefont
			{H.}~\bibnamefont {Takahashi}}, \bibinfo {author} {\bibfnamefont {A.~B.}\
			\bibnamefont {Shick}}, \ and\ \bibinfo {author} {\bibfnamefont
			{T.}~\bibnamefont {Jungwirth}},\ }\href {\doibase 10.1038/nmat2983}
	{\bibfield  {journal} {\bibinfo  {journal} {Nature materials}\ }\textbf
		{\bibinfo {volume} {10}},\ \bibinfo {pages} {347} (\bibinfo {year}
		{2011})}\BibitemShut {NoStop}%
	\bibitem [{\citenamefont {Wang}\ \emph {et~al.}(2012)\citenamefont {Wang},
		\citenamefont {Song}, \citenamefont {Cui}, \citenamefont {Wang},
		\citenamefont {Zeng},\ and\ \citenamefont {Pan}}]{Wang2012a}%
	\BibitemOpen
	\bibfield  {author} {\bibinfo {author} {\bibfnamefont {Y.~Y.}\ \bibnamefont
			{Wang}}, \bibinfo {author} {\bibfnamefont {C.}~\bibnamefont {Song}}, \bibinfo
		{author} {\bibfnamefont {B.}~\bibnamefont {Cui}}, \bibinfo {author}
		{\bibfnamefont {G.~Y.}\ \bibnamefont {Wang}}, \bibinfo {author}
		{\bibfnamefont {F.}~\bibnamefont {Zeng}}, \ and\ \bibinfo {author}
		{\bibfnamefont {F.}~\bibnamefont {Pan}},\ }\href {\doibase
		10.1103/PhysRevLett.109.137201} {\bibfield  {journal} {\bibinfo  {journal}
			{Physical Review Letters}\ }\textbf {\bibinfo {volume} {109}},\ \bibinfo
		{pages} {137201} (\bibinfo {year} {2012})}\BibitemShut {NoStop}%
	\bibitem [{\citenamefont {Marti}\ \emph {et~al.}(2014)\citenamefont {Marti},
		\citenamefont {Fina}, \citenamefont {Frontera}, \citenamefont {Liu},
		\citenamefont {Wadley}, \citenamefont {He}, \citenamefont {Paull},
		\citenamefont {Clarkson}, \citenamefont {Kudrnovsk\'{y}}, \citenamefont
		{Turek}, \citenamefont {Kune\v{s}}, \citenamefont {Yi}, \citenamefont {Chu},
		\citenamefont {Nelson}, \citenamefont {You}, \citenamefont {Arenholz},
		\citenamefont {Salahuddin}, \citenamefont {Fontcuberta}, \citenamefont
		{Jungwirth},\ and\ \citenamefont {Ramesh}}]{Marti2014}%
	\BibitemOpen
	\bibfield  {author} {\bibinfo {author} {\bibfnamefont {X.}~\bibnamefont
			{Marti}}, \bibinfo {author} {\bibfnamefont {I.}~\bibnamefont {Fina}},
		\bibinfo {author} {\bibfnamefont {C.}~\bibnamefont {Frontera}}, \bibinfo
		{author} {\bibfnamefont {J.}~\bibnamefont {Liu}}, \bibinfo {author}
		{\bibfnamefont {P.}~\bibnamefont {Wadley}}, \bibinfo {author} {\bibfnamefont
			{Q.}~\bibnamefont {He}}, \bibinfo {author} {\bibfnamefont {R.~J.}\
			\bibnamefont {Paull}}, \bibinfo {author} {\bibfnamefont {J.~D.}\ \bibnamefont
			{Clarkson}}, \bibinfo {author} {\bibfnamefont {J.}~\bibnamefont
			{Kudrnovsk\'{y}}}, \bibinfo {author} {\bibfnamefont {I.}~\bibnamefont
			{Turek}}, \bibinfo {author} {\bibfnamefont {J.}~\bibnamefont {Kune\v{s}}},
		\bibinfo {author} {\bibfnamefont {D.}~\bibnamefont {Yi}}, \bibinfo {author}
		{\bibfnamefont {J.-H.}\ \bibnamefont {Chu}}, \bibinfo {author} {\bibfnamefont
			{C.~T.}\ \bibnamefont {Nelson}}, \bibinfo {author} {\bibfnamefont
			{L.}~\bibnamefont {You}}, \bibinfo {author} {\bibfnamefont {E.}~\bibnamefont
			{Arenholz}}, \bibinfo {author} {\bibfnamefont {S.}~\bibnamefont
			{Salahuddin}}, \bibinfo {author} {\bibfnamefont {J.}~\bibnamefont
			{Fontcuberta}}, \bibinfo {author} {\bibfnamefont {T.}~\bibnamefont
			{Jungwirth}}, \ and\ \bibinfo {author} {\bibfnamefont {R.}~\bibnamefont
			{Ramesh}},\ }\href {\doibase 10.1038/nmat3861} {\bibfield  {journal}
		{\bibinfo  {journal} {Nature materials}\ }\textbf {\bibinfo {volume} {13}},\
		\bibinfo {pages} {367} (\bibinfo {year} {2014})}\BibitemShut {NoStop}%
	\bibitem [{\citenamefont {Fina}\ \emph {et~al.}(2014)\citenamefont {Fina},
		\citenamefont {Marti}, \citenamefont {Yi}, \citenamefont {Liu}, \citenamefont
		{Chu}, \citenamefont {Rayan-Serrao}, \citenamefont {Suresha}, \citenamefont
		{Shick}, \citenamefont {Zelezn\'{y}}, \citenamefont {Jungwirth},
		\citenamefont {Fontcuberta},\ and\ \citenamefont {Ramesh}}]{Fina2014}%
	\BibitemOpen
	\bibfield  {author} {\bibinfo {author} {\bibfnamefont {I.}~\bibnamefont
			{Fina}}, \bibinfo {author} {\bibfnamefont {X.}~\bibnamefont {Marti}},
		\bibinfo {author} {\bibfnamefont {D.}~\bibnamefont {Yi}}, \bibinfo {author}
		{\bibfnamefont {J.}~\bibnamefont {Liu}}, \bibinfo {author} {\bibfnamefont
			{J.~H.}\ \bibnamefont {Chu}}, \bibinfo {author} {\bibfnamefont
			{C.}~\bibnamefont {Rayan-Serrao}}, \bibinfo {author} {\bibfnamefont
			{S.}~\bibnamefont {Suresha}}, \bibinfo {author} {\bibfnamefont {a.~B.}\
			\bibnamefont {Shick}}, \bibinfo {author} {\bibfnamefont {J.}~\bibnamefont
			{Zelezn\'{y}}}, \bibinfo {author} {\bibfnamefont {T.}~\bibnamefont
			{Jungwirth}}, \bibinfo {author} {\bibfnamefont {J.}~\bibnamefont
			{Fontcuberta}}, \ and\ \bibinfo {author} {\bibfnamefont {R.}~\bibnamefont
			{Ramesh}},\ }\href {\doibase 10.1038/ncomms5671} {\bibfield  {journal}
		{\bibinfo  {journal} {Nature communications}\ }\textbf {\bibinfo {volume}
			{5}},\ \bibinfo {pages} {4671} (\bibinfo {year} {2014})}\BibitemShut
	{NoStop}%
	\bibitem [{\citenamefont {Wadley}\ \emph {et~al.}()\citenamefont {Wadley},
		\citenamefont {Howells}, \citenamefont {Zelezny}, \citenamefont {Andrews},
		\citenamefont {Hills}, \citenamefont {Campion}, \citenamefont {Novak},
		\citenamefont {Freimuth}, \citenamefont {Mokrousov}, \citenamefont
		{Rushforth}, \citenamefont {Edmonds}, \citenamefont {Gallagher},\ and\
		\citenamefont {Jungwirth}}]{Wadley2015}%
	\BibitemOpen
	\bibfield  {author} {\bibinfo {author} {\bibfnamefont {P.}~\bibnamefont
			{Wadley}}, \bibinfo {author} {\bibfnamefont {B.}~\bibnamefont {Howells}},
		\bibinfo {author} {\bibfnamefont {J.}~\bibnamefont {Zelezny}}, \bibinfo
		{author} {\bibfnamefont {C.}~\bibnamefont {Andrews}}, \bibinfo {author}
		{\bibfnamefont {V.}~\bibnamefont {Hills}}, \bibinfo {author} {\bibfnamefont
			{R.~P.}\ \bibnamefont {Campion}}, \bibinfo {author} {\bibfnamefont
			{V.}~\bibnamefont {Novak}}, \bibinfo {author} {\bibfnamefont
			{F.}~\bibnamefont {Freimuth}}, \bibinfo {author} {\bibfnamefont
			{Y.}~\bibnamefont {Mokrousov}}, \bibinfo {author} {\bibfnamefont {A.~W.}\
			\bibnamefont {Rushforth}}, \bibinfo {author} {\bibfnamefont {K.~W.}\
			\bibnamefont {Edmonds}}, \bibinfo {author} {\bibfnamefont {B.~L.}\
			\bibnamefont {Gallagher}}, \ and\ \bibinfo {author} {\bibfnamefont
			{T.}~\bibnamefont {Jungwirth}},\ }\href {http://arxiv.org/abs/1503.03765} {\
	}\Eprint {http://arxiv.org/abs/1503.03765} {arXiv:1503.03765} \BibitemShut
	{NoStop}%
	\bibitem [{\citenamefont {Wang}\ \emph {et~al.}(2014)\citenamefont {Wang},
		\citenamefont {Du}, \citenamefont {Hammel},\ and\ \citenamefont
		{Yang}}]{Wang2014d}%
	\BibitemOpen
	\bibfield  {author} {\bibinfo {author} {\bibfnamefont {H.}~\bibnamefont
			{Wang}}, \bibinfo {author} {\bibfnamefont {C.}~\bibnamefont {Du}}, \bibinfo
		{author} {\bibfnamefont {P.~C.}\ \bibnamefont {Hammel}}, \ and\ \bibinfo
		{author} {\bibfnamefont {F.}~\bibnamefont {Yang}},\ }\href {\doibase
		10.1103/PhysRevLett.113.097202} {\bibfield  {journal} {\bibinfo  {journal}
			{Physical Review Letters}\ }\textbf {\bibinfo {volume} {113}},\ \bibinfo
		{pages} {097202} (\bibinfo {year} {2014})}\BibitemShut {NoStop}%
	\bibitem [{\citenamefont {Hahn}\ \emph {et~al.}(2014)\citenamefont {Hahn},
		\citenamefont {de~Loubens}, \citenamefont {Naletov}, \citenamefont {{Ben
				Youssef}}, \citenamefont {Klein},\ and\ \citenamefont {Viret}}]{Hahn2014}%
	\BibitemOpen
	\bibfield  {author} {\bibinfo {author} {\bibfnamefont {C.}~\bibnamefont
			{Hahn}}, \bibinfo {author} {\bibfnamefont {G.}~\bibnamefont {de~Loubens}},
		\bibinfo {author} {\bibfnamefont {V.~V.}\ \bibnamefont {Naletov}}, \bibinfo
		{author} {\bibfnamefont {J.}~\bibnamefont {{Ben Youssef}}}, \bibinfo {author}
		{\bibfnamefont {O.}~\bibnamefont {Klein}}, \ and\ \bibinfo {author}
		{\bibfnamefont {M.}~\bibnamefont {Viret}},\ }\href {\doibase
		10.1209/0295-5075/108/57005} {\bibfield  {journal} {\bibinfo  {journal} {EPL
				(Europhysics Letters)}\ }\textbf {\bibinfo {volume} {108}},\ \bibinfo {pages}
		{57005} (\bibinfo {year} {2014})}\BibitemShut {NoStop}%
	\bibitem [{\citenamefont {Moriyama}\ \emph {et~al.}()\citenamefont {Moriyama},
		\citenamefont {Nagata}, \citenamefont {Tanaka}, \citenamefont {Kim},
		\citenamefont {Almasi},\ and\ \citenamefont {Wang}}]{Moriyama}%
	\BibitemOpen
	\bibfield  {author} {\bibinfo {author} {\bibfnamefont {T.}~\bibnamefont
			{Moriyama}}, \bibinfo {author} {\bibfnamefont {M.}~\bibnamefont {Nagata}},
		\bibinfo {author} {\bibfnamefont {K.}~\bibnamefont {Tanaka}}, \bibinfo
		{author} {\bibfnamefont {K.-j.}\ \bibnamefont {Kim}}, \bibinfo {author}
		{\bibfnamefont {H.}~\bibnamefont {Almasi}}, \ and\ \bibinfo {author}
		{\bibfnamefont {W.~G.}\ \bibnamefont {Wang}},\ }\href@noop {} {\bibinfo
		{journal} {arXiv:1411.4100}\ }\BibitemShut {NoStop}%
	\bibitem [{\citenamefont {Ralph}\ and\ \citenamefont
		{Stiles}(2008)}]{Ralph2008}%
	\BibitemOpen
	\bibfield  {journal} {  }\bibfield  {author} {\bibinfo {author} {\bibfnamefont
			{D.}~\bibnamefont {Ralph}}\ and\ \bibinfo {author} {\bibfnamefont
			{M.}~\bibnamefont {Stiles}},\ }\href {\doibase 10.1016/j.jmmm.2007.12.019}
	{\bibfield  {journal} {\bibinfo  {journal} {Journal of Magnetism and Magnetic
				Materials}\ }\textbf {\bibinfo {volume} {320}},\ \bibinfo {pages} {1190}
		(\bibinfo {year} {2008})}\BibitemShut {NoStop}%
	\bibitem [{\citenamefont {Mendes}\ \emph {et~al.}(2014)\citenamefont {Mendes},
		\citenamefont {Cunha}, \citenamefont {{Alves Santos}}, \citenamefont
		{Ribeiro}, \citenamefont {Machado}, \citenamefont
		{Rodr\'{\i}guez-Su\'{a}rez}, \citenamefont {Azevedo},\ and\ \citenamefont
		{Rezende}}]{Mendes2014}%
	\BibitemOpen
	\bibfield  {author} {\bibinfo {author} {\bibfnamefont {J.~B.~S.}\
			\bibnamefont {Mendes}}, \bibinfo {author} {\bibfnamefont {R.~O.}\
			\bibnamefont {Cunha}}, \bibinfo {author} {\bibfnamefont {O.}~\bibnamefont
			{{Alves Santos}}}, \bibinfo {author} {\bibfnamefont {P.~R.~T.}\ \bibnamefont
			{Ribeiro}}, \bibinfo {author} {\bibfnamefont {F.~L.~A.}\ \bibnamefont
			{Machado}}, \bibinfo {author} {\bibfnamefont {R.~L.}\ \bibnamefont
			{Rodr\'{\i}guez-Su\'{a}rez}}, \bibinfo {author} {\bibfnamefont
			{A.}~\bibnamefont {Azevedo}}, \ and\ \bibinfo {author} {\bibfnamefont
			{S.~M.}\ \bibnamefont {Rezende}},\ }\href {\doibase
		10.1103/PhysRevB.89.140406} {\bibfield  {journal} {\bibinfo  {journal}
			{Physical Review B}\ }\textbf {\bibinfo {volume} {89}},\ \bibinfo {pages}
		{140406} (\bibinfo {year} {2014})}\BibitemShut {NoStop}%
	\bibitem [{\citenamefont {Zhang}\ \emph {et~al.}(2014)\citenamefont {Zhang},
		\citenamefont {Jungfleisch}, \citenamefont {Jiang}, \citenamefont {Pearson},
		\citenamefont {Hoffmann}, \citenamefont {Freimuth},\ and\ \citenamefont
		{Mokrousov}}]{Zhang2014}%
	\BibitemOpen
	\bibfield  {author} {\bibinfo {author} {\bibfnamefont {W.}~\bibnamefont
			{Zhang}}, \bibinfo {author} {\bibfnamefont {M.~B.}\ \bibnamefont
			{Jungfleisch}}, \bibinfo {author} {\bibfnamefont {W.}~\bibnamefont {Jiang}},
		\bibinfo {author} {\bibfnamefont {J.~E.}\ \bibnamefont {Pearson}}, \bibinfo
		{author} {\bibfnamefont {A.}~\bibnamefont {Hoffmann}}, \bibinfo {author}
		{\bibfnamefont {F.}~\bibnamefont {Freimuth}}, \ and\ \bibinfo {author}
		{\bibfnamefont {Y.}~\bibnamefont {Mokrousov}},\ }\href {\doibase
		10.1103/PhysRevLett.113.196602} {\bibfield  {journal} {\bibinfo  {journal}
			{Physical Review Letters}\ }\textbf {\bibinfo {volume} {113}},\ \bibinfo
		{pages} {196602} (\bibinfo {year} {2014})}\BibitemShut {NoStop}%
	\bibitem [{\citenamefont {Bernevig}\ and\ \citenamefont
		{Vafek}(2005)}]{Bernevig2005c}%
	\BibitemOpen
	\bibfield  {author} {\bibinfo {author} {\bibfnamefont {B.}~\bibnamefont
			{Bernevig}}\ and\ \bibinfo {author} {\bibfnamefont {O.}~\bibnamefont
			{Vafek}},\ }\href {\doibase 10.1103/PhysRevB.72.033203} {\bibfield  {journal}
		{\bibinfo  {journal} {Physical Review B}\ }\textbf {\bibinfo {volume} {72}},\
		\bibinfo {pages} {033203} (\bibinfo {year} {2005})}\BibitemShut {NoStop}%
	\bibitem [{\citenamefont {Manchon}\ and\ \citenamefont
		{Zhang}(2008)}]{Manchon2008}%
	\BibitemOpen
	\bibfield  {author} {\bibinfo {author} {\bibfnamefont {A.}~\bibnamefont
			{Manchon}}\ and\ \bibinfo {author} {\bibfnamefont {S.}~\bibnamefont
			{Zhang}},\ }\href {\doibase 10.1103/PhysRevB.78.212405} {\bibfield  {journal}
		{\bibinfo  {journal} {Physical Review B}\ }\textbf {\bibinfo {volume} {78}},\
		\bibinfo {pages} {212405} (\bibinfo {year} {2008})}\BibitemShut {NoStop}%
	\bibitem [{\citenamefont {Chernyshov}\ \emph {et~al.}(2009)\citenamefont
		{Chernyshov}, \citenamefont {Overby}, \citenamefont {Liu}, \citenamefont
		{Furdyna}, \citenamefont {Lyanda-Geller},\ and\ \citenamefont
		{Rokhinson}}]{Chernyshov2009}%
	\BibitemOpen
	\bibfield  {author} {\bibinfo {author} {\bibfnamefont {A.}~\bibnamefont
			{Chernyshov}}, \bibinfo {author} {\bibfnamefont {M.}~\bibnamefont {Overby}},
		\bibinfo {author} {\bibfnamefont {X.}~\bibnamefont {Liu}}, \bibinfo {author}
		{\bibfnamefont {J.~K.}\ \bibnamefont {Furdyna}}, \bibinfo {author}
		{\bibfnamefont {Y.}~\bibnamefont {Lyanda-Geller}}, \ and\ \bibinfo {author}
		{\bibfnamefont {L.~P.}\ \bibnamefont {Rokhinson}},\ }\href {\doibase
		10.1038/nphys1362} {\bibfield  {journal} {\bibinfo  {journal} {Nature
				Physics}\ }\textbf {\bibinfo {volume} {5}},\ \bibinfo {pages} {656} (\bibinfo
		{year} {2009})}\BibitemShut {NoStop}%
	\bibitem [{\citenamefont {Fang}\ \emph {et~al.}(2011)\citenamefont {Fang},
		\citenamefont {Kurebayashi}, \citenamefont {Wunderlich}, \citenamefont
		{V\'{y}born\'{y}}, \citenamefont {Z\^{a}rbo}, \citenamefont {Campion},
		\citenamefont {Casiraghi}, \citenamefont {Gallagher}, \citenamefont
		{Jungwirth},\ and\ \citenamefont {Ferguson}}]{Fang2011}%
	\BibitemOpen
	\bibfield  {author} {\bibinfo {author} {\bibfnamefont {D.}~\bibnamefont
			{Fang}}, \bibinfo {author} {\bibfnamefont {H.}~\bibnamefont {Kurebayashi}},
		\bibinfo {author} {\bibfnamefont {J.}~\bibnamefont {Wunderlich}}, \bibinfo
		{author} {\bibfnamefont {K.}~\bibnamefont {V\'{y}born\'{y}}}, \bibinfo
		{author} {\bibfnamefont {L.~P.}\ \bibnamefont {Z\^{a}rbo}}, \bibinfo {author}
		{\bibfnamefont {R.~P.}\ \bibnamefont {Campion}}, \bibinfo {author}
		{\bibfnamefont {A.}~\bibnamefont {Casiraghi}}, \bibinfo {author}
		{\bibfnamefont {B.~L.}\ \bibnamefont {Gallagher}}, \bibinfo {author}
		{\bibfnamefont {T.}~\bibnamefont {Jungwirth}}, \ and\ \bibinfo {author}
		{\bibfnamefont {A.~J.}\ \bibnamefont {Ferguson}},\ }\href {\doibase
		10.1038/nnano.2011.68} {\bibfield  {journal} {\bibinfo  {journal} {Nature
				Nanotechnology}\ }\textbf {\bibinfo {volume} {6}},\ \bibinfo {pages} {413}
		(\bibinfo {year} {2011})}\BibitemShut {NoStop}%
	\bibitem [{\citenamefont {Miron}\ \emph {et~al.}(2010)\citenamefont {Miron},
		\citenamefont {Gaudin}, \citenamefont {Auffret}, \citenamefont {Rodmacq},
		\citenamefont {Schuhl}, \citenamefont {Pizzini}, \citenamefont {Vogel},\ and\
		\citenamefont {Gambardella}}]{Miron2010}%
	\BibitemOpen
	\bibfield  {author} {\bibinfo {author} {\bibfnamefont {I.~M.}\ \bibnamefont
			{Miron}}, \bibinfo {author} {\bibfnamefont {G.}~\bibnamefont {Gaudin}},
		\bibinfo {author} {\bibfnamefont {S.}~\bibnamefont {Auffret}}, \bibinfo
		{author} {\bibfnamefont {B.}~\bibnamefont {Rodmacq}}, \bibinfo {author}
		{\bibfnamefont {A.}~\bibnamefont {Schuhl}}, \bibinfo {author} {\bibfnamefont
			{S.}~\bibnamefont {Pizzini}}, \bibinfo {author} {\bibfnamefont
			{J.}~\bibnamefont {Vogel}}, \ and\ \bibinfo {author} {\bibfnamefont
			{P.}~\bibnamefont {Gambardella}},\ }\href {\doibase 10.1038/nmat2613}
	{\bibfield  {journal} {\bibinfo  {journal} {Nature Materials}\ }\textbf
		{\bibinfo {volume} {9}},\ \bibinfo {pages} {230} (\bibinfo {year}
		{2010})}\BibitemShut {NoStop}%
	\bibitem [{\citenamefont {Miron}\ \emph {et~al.}(2011)\citenamefont {Miron},
		\citenamefont {Garello}, \citenamefont {Gaudin}, \citenamefont {Zermatten},
		\citenamefont {Costache}, \citenamefont {Auffret}, \citenamefont {Bandiera},
		\citenamefont {Rodmacq}, \citenamefont {Schuhl},\ and\ \citenamefont
		{Gambardella}}]{Miron2011}%
	\BibitemOpen
	\bibfield  {author} {\bibinfo {author} {\bibfnamefont {I.~M.}\ \bibnamefont
			{Miron}}, \bibinfo {author} {\bibfnamefont {K.}~\bibnamefont {Garello}},
		\bibinfo {author} {\bibfnamefont {G.}~\bibnamefont {Gaudin}}, \bibinfo
		{author} {\bibfnamefont {P.-J.}\ \bibnamefont {Zermatten}}, \bibinfo {author}
		{\bibfnamefont {M.~V.}\ \bibnamefont {Costache}}, \bibinfo {author}
		{\bibfnamefont {S.}~\bibnamefont {Auffret}}, \bibinfo {author} {\bibfnamefont
			{S.}~\bibnamefont {Bandiera}}, \bibinfo {author} {\bibfnamefont
			{B.}~\bibnamefont {Rodmacq}}, \bibinfo {author} {\bibfnamefont
			{A.}~\bibnamefont {Schuhl}}, \ and\ \bibinfo {author} {\bibfnamefont
			{P.}~\bibnamefont {Gambardella}},\ }\href {\doibase 10.1038/nature10309}
	{\bibfield  {journal} {\bibinfo  {journal} {Nature}\ }\textbf {\bibinfo
			{volume} {476}},\ \bibinfo {pages} {189} (\bibinfo {year}
		{2011})}\BibitemShut {NoStop}%
	\bibitem [{\citenamefont {Liu}\ \emph {et~al.}(2012)\citenamefont {Liu},
		\citenamefont {Pai}, \citenamefont {Li}, \citenamefont {Tseng}, \citenamefont
		{Ralph},\ and\ \citenamefont {Buhrman}}]{Liu2012}%
	\BibitemOpen
	\bibfield  {author} {\bibinfo {author} {\bibfnamefont {L.}~\bibnamefont
			{Liu}}, \bibinfo {author} {\bibfnamefont {C.-F.}\ \bibnamefont {Pai}},
		\bibinfo {author} {\bibfnamefont {Y.}~\bibnamefont {Li}}, \bibinfo {author}
		{\bibfnamefont {H.~W.}\ \bibnamefont {Tseng}}, \bibinfo {author}
		{\bibfnamefont {D.~C.}\ \bibnamefont {Ralph}}, \ and\ \bibinfo {author}
		{\bibfnamefont {R.~A.}\ \bibnamefont {Buhrman}},\ }\href {\doibase
		10.1126/science.1218197} {\bibfield  {journal} {\bibinfo  {journal} {Science
				(New York, N.Y.)}\ }\textbf {\bibinfo {volume} {336}},\ \bibinfo {pages}
		{555} (\bibinfo {year} {2012})}\BibitemShut {NoStop}%
	\bibitem [{\citenamefont {Garello}\ \emph {et~al.}(2013)\citenamefont
		{Garello}, \citenamefont {Miron}, \citenamefont {Avci}, \citenamefont
		{Freimuth}, \citenamefont {Mokrousov}, \citenamefont {Bl\"{u}gel},
		\citenamefont {Auffret}, \citenamefont {Boulle}, \citenamefont {Gaudin},\
		and\ \citenamefont {Gambardella}}]{Garello2013}%
	\BibitemOpen
	\bibfield  {author} {\bibinfo {author} {\bibfnamefont {K.}~\bibnamefont
			{Garello}}, \bibinfo {author} {\bibfnamefont {I.~M.}\ \bibnamefont {Miron}},
		\bibinfo {author} {\bibfnamefont {C.~O.}\ \bibnamefont {Avci}}, \bibinfo
		{author} {\bibfnamefont {F.}~\bibnamefont {Freimuth}}, \bibinfo {author}
		{\bibfnamefont {Y.}~\bibnamefont {Mokrousov}}, \bibinfo {author}
		{\bibfnamefont {S.}~\bibnamefont {Bl\"{u}gel}}, \bibinfo {author}
		{\bibfnamefont {S.}~\bibnamefont {Auffret}}, \bibinfo {author} {\bibfnamefont
			{O.}~\bibnamefont {Boulle}}, \bibinfo {author} {\bibfnamefont
			{G.}~\bibnamefont {Gaudin}}, \ and\ \bibinfo {author} {\bibfnamefont
			{P.}~\bibnamefont {Gambardella}},\ }\href {\doibase 10.1038/nnano.2013.145}
	{\bibfield  {journal} {\bibinfo  {journal} {Nature Nanotechnology}\ }\textbf
		{\bibinfo {volume} {8}},\ \bibinfo {pages} {587} (\bibinfo {year}
		{2013})}\BibitemShut {NoStop}%
	\bibitem [{\citenamefont {Kurebayashi}\ \emph {et~al.}(2014)\citenamefont
		{Kurebayashi}, \citenamefont {Sinova}, \citenamefont {Fang}, \citenamefont
		{Irvine}, \citenamefont {Skinner}, \citenamefont {Wunderlich}, \citenamefont
		{Nov\'{a}k}, \citenamefont {Campion}, \citenamefont {Gallagher},
		\citenamefont {Vehstedt}, \citenamefont {Z\^{a}rbo}, \citenamefont
		{V\'{y}born\'{y}}, \citenamefont {Ferguson},\ and\ \citenamefont
		{Jungwirth}}]{Kurebayashi2014}%
	\BibitemOpen
	\bibfield  {author} {\bibinfo {author} {\bibfnamefont {H.}~\bibnamefont
			{Kurebayashi}}, \bibinfo {author} {\bibfnamefont {J.}~\bibnamefont {Sinova}},
		\bibinfo {author} {\bibfnamefont {D.}~\bibnamefont {Fang}}, \bibinfo {author}
		{\bibfnamefont {A.~C.}\ \bibnamefont {Irvine}}, \bibinfo {author}
		{\bibfnamefont {T.~D.}\ \bibnamefont {Skinner}}, \bibinfo {author}
		{\bibfnamefont {J.}~\bibnamefont {Wunderlich}}, \bibinfo {author}
		{\bibfnamefont {V.}~\bibnamefont {Nov\'{a}k}}, \bibinfo {author}
		{\bibfnamefont {R.~P.}\ \bibnamefont {Campion}}, \bibinfo {author}
		{\bibfnamefont {B.~L.}\ \bibnamefont {Gallagher}}, \bibinfo {author}
		{\bibfnamefont {E.~K.}\ \bibnamefont {Vehstedt}}, \bibinfo {author}
		{\bibfnamefont {L.~P.}\ \bibnamefont {Z\^{a}rbo}}, \bibinfo {author}
		{\bibfnamefont {K.}~\bibnamefont {V\'{y}born\'{y}}}, \bibinfo {author}
		{\bibfnamefont {A.~J.}\ \bibnamefont {Ferguson}}, \ and\ \bibinfo {author}
		{\bibfnamefont {T.}~\bibnamefont {Jungwirth}},\ }\href {\doibase
		10.1038/nnano.2014.15} {\bibfield  {journal} {\bibinfo  {journal} {Nature
				Nanotechnology}\ }\textbf {\bibinfo {volume} {9}},\ \bibinfo {pages} {211}
		(\bibinfo {year} {2014})}\BibitemShut {NoStop}%
	\bibitem [{\citenamefont {Costache}\ \emph {et~al.}(2006)\citenamefont
		{Costache}, \citenamefont {Watts}, \citenamefont {Sladkov}, \citenamefont
		{van~der Wal},\ and\ \citenamefont {van Wees}}]{Costache2006}%
	\BibitemOpen
	\bibfield  {author} {\bibinfo {author} {\bibfnamefont {M.~V.}\ \bibnamefont
			{Costache}}, \bibinfo {author} {\bibfnamefont {S.~M.}\ \bibnamefont {Watts}},
		\bibinfo {author} {\bibfnamefont {M.}~\bibnamefont {Sladkov}}, \bibinfo
		{author} {\bibfnamefont {C.~H.}\ \bibnamefont {van~der Wal}}, \ and\ \bibinfo
		{author} {\bibfnamefont {B.~J.}\ \bibnamefont {van Wees}},\ }\href {\doibase
		10.1063/1.2400058} {\bibfield  {journal} {\bibinfo  {journal} {Applied
				Physics Letters}\ }\textbf {\bibinfo {volume} {89}},\ \bibinfo {pages}
		{232115} (\bibinfo {year} {2006})}\BibitemShut {NoStop}%
	\bibitem [{\citenamefont {Yakata}\ \emph {et~al.}(2006)\citenamefont {Yakata},
		\citenamefont {Ando}, \citenamefont {Miyazaki},\ and\ \citenamefont
		{Mizukami}}]{Yakata2006}%
	\BibitemOpen
	\bibfield  {author} {\bibinfo {author} {\bibfnamefont {S.}~\bibnamefont
			{Yakata}}, \bibinfo {author} {\bibfnamefont {Y.}~\bibnamefont {Ando}},
		\bibinfo {author} {\bibfnamefont {T.}~\bibnamefont {Miyazaki}}, \ and\
		\bibinfo {author} {\bibfnamefont {S.}~\bibnamefont {Mizukami}},\ }\href
	{\doibase 10.1143/JJAP.45.3892} {\bibfield  {journal} {\bibinfo  {journal}
			{Japanese Journal of Applied Physics}\ }\textbf {\bibinfo {volume} {45}},\
		\bibinfo {pages} {3892} (\bibinfo {year} {2006})}\BibitemShut {NoStop}%
	\bibitem [{\citenamefont {Liu}\ \emph {et~al.}(2011)\citenamefont {Liu},
		\citenamefont {Moriyama}, \citenamefont {Ralph},\ and\ \citenamefont
		{Buhrman}}]{Liu2011}%
	\BibitemOpen
	\bibfield  {author} {\bibinfo {author} {\bibfnamefont {L.}~\bibnamefont
			{Liu}}, \bibinfo {author} {\bibfnamefont {T.}~\bibnamefont {Moriyama}},
		\bibinfo {author} {\bibfnamefont {D.}~\bibnamefont {Ralph}}, \ and\ \bibinfo
		{author} {\bibfnamefont {R.}~\bibnamefont {Buhrman}},\ }\href {\doibase
		10.1103/PhysRevLett.106.036601} {\bibfield  {journal} {\bibinfo  {journal}
			{Physical Review Letters}\ }\textbf {\bibinfo {volume} {106}},\ \bibinfo
		{pages} {1} (\bibinfo {year} {2011})}\BibitemShut {NoStop}%
	\bibitem [{\citenamefont {Tanaka}\ \emph {et~al.}(2008)\citenamefont {Tanaka},
		\citenamefont {Kontani}, \citenamefont {Naito}, \citenamefont {Naito},
		\citenamefont {Hirashima}, \citenamefont {Yamada},\ and\ \citenamefont
		{Inoue}}]{Tanaka2008}%
	\BibitemOpen
	\bibfield  {author} {\bibinfo {author} {\bibfnamefont {T.}~\bibnamefont
			{Tanaka}}, \bibinfo {author} {\bibfnamefont {H.}~\bibnamefont {Kontani}},
		\bibinfo {author} {\bibfnamefont {M.}~\bibnamefont {Naito}}, \bibinfo
		{author} {\bibfnamefont {T.}~\bibnamefont {Naito}}, \bibinfo {author}
		{\bibfnamefont {D.}~\bibnamefont {Hirashima}}, \bibinfo {author}
		{\bibfnamefont {K.}~\bibnamefont {Yamada}}, \ and\ \bibinfo {author}
		{\bibfnamefont {J.}~\bibnamefont {Inoue}},\ }\href {\doibase
		10.1103/PhysRevB.77.165117} {\bibfield  {journal} {\bibinfo  {journal}
			{Physical Review B}\ }\textbf {\bibinfo {volume} {77}},\ \bibinfo {pages}
		{165117} (\bibinfo {year} {2008})}\BibitemShut {NoStop}%
	\bibitem [{\citenamefont {Acharyya}\ \emph {et~al.}(2011)\citenamefont
		{Acharyya}, \citenamefont {Nguyen}, \citenamefont {Pratt},\ and\
		\citenamefont {Bass}}]{Acharyya2011}%
	\BibitemOpen
	\bibfield  {author} {\bibinfo {author} {\bibfnamefont {R.}~\bibnamefont
			{Acharyya}}, \bibinfo {author} {\bibfnamefont {H.~Y.~T.}\ \bibnamefont
			{Nguyen}}, \bibinfo {author} {\bibfnamefont {W.~P.}\ \bibnamefont {Pratt}}, \
		and\ \bibinfo {author} {\bibfnamefont {J.}~\bibnamefont {Bass}},\ }\href
	{\doibase 10.1063/1.3535340} {\bibfield  {journal} {\bibinfo  {journal}
			{Journal of Applied Physics}\ }\textbf {\bibinfo {volume} {109}},\ \bibinfo
		{pages} {07C503} (\bibinfo {year} {2011})}\BibitemShut {NoStop}%
	\bibitem [{\citenamefont {Stiles}\ and\ \citenamefont
		{McMichael}(1999)}]{Stiles1999}%
	\BibitemOpen
	\bibfield  {author} {\bibinfo {author} {\bibfnamefont {M.}~\bibnamefont
			{Stiles}}\ and\ \bibinfo {author} {\bibfnamefont {R.}~\bibnamefont
			{McMichael}},\ }\href {\doibase 10.1103/PhysRevB.59.3722} {\bibfield
		{journal} {\bibinfo  {journal} {Physical Review B}\ }\textbf {\bibinfo
			{volume} {59}},\ \bibinfo {pages} {3722} (\bibinfo {year}
		{1999})}\BibitemShut {NoStop}%
	\bibitem [{\citenamefont {Stiles}\ and\ \citenamefont
		{McMichael}(2001)}]{Stiles2001}%
	\BibitemOpen
	\bibfield  {author} {\bibinfo {author} {\bibfnamefont {M.}~\bibnamefont
			{Stiles}}\ and\ \bibinfo {author} {\bibfnamefont {R.}~\bibnamefont
			{McMichael}},\ }\href {\doibase 10.1103/PhysRevB.63.064405} {\bibfield
		{journal} {\bibinfo  {journal} {Physical Review B}\ }\textbf {\bibinfo
			{volume} {63}},\ \bibinfo {pages} {064405} (\bibinfo {year}
		{2001})}\BibitemShut {NoStop}%
	\bibitem [{\citenamefont {Ali}\ \emph {et~al.}(2003)\citenamefont {Ali},
		\citenamefont {Marrows}, \citenamefont {Al-Jawad}, \citenamefont {Hickey},
		\citenamefont {Misra}, \citenamefont {Nowak},\ and\ \citenamefont
		{Usadel}}]{Ali2003}%
	\BibitemOpen
	\bibfield  {author} {\bibinfo {author} {\bibfnamefont {M.}~\bibnamefont
			{Ali}}, \bibinfo {author} {\bibfnamefont {C.}~\bibnamefont {Marrows}},
		\bibinfo {author} {\bibfnamefont {M.}~\bibnamefont {Al-Jawad}}, \bibinfo
		{author} {\bibfnamefont {B.}~\bibnamefont {Hickey}}, \bibinfo {author}
		{\bibfnamefont {a.}~\bibnamefont {Misra}}, \bibinfo {author} {\bibfnamefont
			{U.}~\bibnamefont {Nowak}}, \ and\ \bibinfo {author} {\bibfnamefont
			{K.}~\bibnamefont {Usadel}},\ }\href {\doibase 10.1103/PhysRevB.68.214420}
	{\bibfield  {journal} {\bibinfo  {journal} {Physical Review B}\ }\textbf
		{\bibinfo {volume} {68}},\ \bibinfo {pages} {214420} (\bibinfo {year}
		{2003})}\BibitemShut {NoStop}%
	\bibitem [{\citenamefont {McCord}\ \emph {et~al.}(2004)\citenamefont {McCord},
		\citenamefont {Mattheis},\ and\ \citenamefont {Elefant}}]{McCord2004}%
	\BibitemOpen
	\bibfield  {author} {\bibinfo {author} {\bibfnamefont {J.}~\bibnamefont
			{McCord}}, \bibinfo {author} {\bibfnamefont {R.}~\bibnamefont {Mattheis}}, \
		and\ \bibinfo {author} {\bibfnamefont {D.}~\bibnamefont {Elefant}},\ }\href
	{\doibase 10.1103/PhysRevB.70.094420} {\bibfield  {journal} {\bibinfo
			{journal} {Physical Review B}\ }\textbf {\bibinfo {volume} {70}},\ \bibinfo
		{pages} {094420} (\bibinfo {year} {2004})}\BibitemShut {NoStop}%
	\bibitem [{\citenamefont {Kim}\ \emph {et~al.}(2014)\citenamefont {Kim},
		\citenamefont {Sinha}, \citenamefont {Mitani}, \citenamefont {Hayashi},
		\citenamefont {Takahashi}, \citenamefont {Maekawa}, \citenamefont
		{Yamanouchi},\ and\ \citenamefont {Ohno}}]{Kim2014}%
	\BibitemOpen
	\bibfield  {author} {\bibinfo {author} {\bibfnamefont {J.}~\bibnamefont
			{Kim}}, \bibinfo {author} {\bibfnamefont {J.}~\bibnamefont {Sinha}}, \bibinfo
		{author} {\bibfnamefont {S.}~\bibnamefont {Mitani}}, \bibinfo {author}
		{\bibfnamefont {M.}~\bibnamefont {Hayashi}}, \bibinfo {author} {\bibfnamefont
			{S.}~\bibnamefont {Takahashi}}, \bibinfo {author} {\bibfnamefont
			{S.}~\bibnamefont {Maekawa}}, \bibinfo {author} {\bibfnamefont
			{M.}~\bibnamefont {Yamanouchi}}, \ and\ \bibinfo {author} {\bibfnamefont
			{H.}~\bibnamefont {Ohno}},\ }\href {\doibase 10.1103/PhysRevB.89.174424}
	{\bibfield  {journal} {\bibinfo  {journal} {Physical Review B - Condensed
				Matter and Materials Physics}\ }\textbf {\bibinfo {volume} {89}},\ \bibinfo
		{pages} {1} (\bibinfo {year} {2014})},\ \Eprint
	{http://arxiv.org/abs/1402.6388} {arXiv:1402.6388} \BibitemShut {NoStop}%
	\bibitem [{\citenamefont {Tserkovnyak}\ \emph {et~al.}(2002)\citenamefont
		{Tserkovnyak}, \citenamefont {Brataas},\ and\ \citenamefont
		{Bauer}}]{Tserkovnyak2002}%
	\BibitemOpen
	\bibfield  {author} {\bibinfo {author} {\bibfnamefont {Y.}~\bibnamefont
			{Tserkovnyak}}, \bibinfo {author} {\bibfnamefont {A.}~\bibnamefont
			{Brataas}}, \ and\ \bibinfo {author} {\bibfnamefont {G.~E.~W.}\ \bibnamefont
			{Bauer}},\ }\href@noop {} {\bibfield  {journal} {\bibinfo  {journal}
			{Physical Review B}\ }\textbf {\bibinfo {volume} {66}},\ \bibinfo {pages}
		{224403} (\bibinfo {year} {2002})}\BibitemShut {NoStop}%
	\bibitem [{\citenamefont {Rezende}\ \emph {et~al.}(2001)\citenamefont
		{Rezende}, \citenamefont {Azevedo}, \citenamefont {Lucena},\ and\
		\citenamefont {de~Aguiar}}]{Rezende2001}%
	\BibitemOpen
	\bibfield  {author} {\bibinfo {author} {\bibfnamefont {S.}~\bibnamefont
			{Rezende}}, \bibinfo {author} {\bibfnamefont {A.}~\bibnamefont {Azevedo}},
		\bibinfo {author} {\bibfnamefont {M.}~\bibnamefont {Lucena}}, \ and\ \bibinfo
		{author} {\bibfnamefont {F.}~\bibnamefont {de~Aguiar}},\ }\href {\doibase
		10.1103/PhysRevB.63.214418} {\bibfield  {journal} {\bibinfo  {journal}
			{Physical Review B}\ }\textbf {\bibinfo {volume} {63}},\ \bibinfo {pages}
		{214418} (\bibinfo {year} {2001})}\BibitemShut {NoStop}%
	\bibitem [{\citenamefont {Yuan}\ \emph {et~al.}(2009)\citenamefont {Yuan},
		\citenamefont {Kang}, \citenamefont {Yu}, \citenamefont {Cao},\ and\
		\citenamefont {Zhao}}]{Yuan2009}%
	\BibitemOpen
	\bibfield  {author} {\bibinfo {author} {\bibfnamefont {S.}~\bibnamefont
			{Yuan}}, \bibinfo {author} {\bibfnamefont {B.}~\bibnamefont {Kang}}, \bibinfo
		{author} {\bibfnamefont {L.}~\bibnamefont {Yu}}, \bibinfo {author}
		{\bibfnamefont {S.}~\bibnamefont {Cao}}, \ and\ \bibinfo {author}
		{\bibfnamefont {X.}~\bibnamefont {Zhao}},\ }\href {\doibase
		10.1063/1.3086292} {\bibfield  {journal} {\bibinfo  {journal} {Journal of
				Applied Physics}\ }\textbf {\bibinfo {volume} {105}},\ \bibinfo {pages}
		{063902} (\bibinfo {year} {2009})}\BibitemShut {NoStop}%
	\bibitem [{\citenamefont {Wei}\ \emph {et~al.}(2007)\citenamefont {Wei},
		\citenamefont {Sharma}, \citenamefont {Nunez}, \citenamefont {Haney},
		\citenamefont {Duine}, \citenamefont {Bass}, \citenamefont {MacDonald},\ and\
		\citenamefont {Tsoi}}]{Wei2007}%
	\BibitemOpen
	\bibfield  {author} {\bibinfo {author} {\bibfnamefont {Z.}~\bibnamefont
			{Wei}}, \bibinfo {author} {\bibfnamefont {A.}~\bibnamefont {Sharma}},
		\bibinfo {author} {\bibfnamefont {A.}~\bibnamefont {Nunez}}, \bibinfo
		{author} {\bibfnamefont {P.}~\bibnamefont {Haney}}, \bibinfo {author}
		{\bibfnamefont {R.}~\bibnamefont {Duine}}, \bibinfo {author} {\bibfnamefont
			{J.}~\bibnamefont {Bass}}, \bibinfo {author} {\bibfnamefont {A.}~\bibnamefont
			{MacDonald}}, \ and\ \bibinfo {author} {\bibfnamefont {M.}~\bibnamefont
			{Tsoi}},\ }\href {\doibase 10.1103/PhysRevLett.98.116603} {\bibfield
		{journal} {\bibinfo  {journal} {Physical Review Letters}\ }\textbf {\bibinfo
			{volume} {98}},\ \bibinfo {pages} {1} (\bibinfo {year} {2007})}\BibitemShut
	{NoStop}%
	\bibitem [{\citenamefont {Urazhdin}\ and\ \citenamefont
		{Anthony}(2007)}]{Urazhdin2007}%
	\BibitemOpen
	\bibfield  {author} {\bibinfo {author} {\bibfnamefont {S.}~\bibnamefont
			{Urazhdin}}\ and\ \bibinfo {author} {\bibfnamefont {N.}~\bibnamefont
			{Anthony}},\ }\href {\doibase 10.1103/PhysRevLett.99.046602} {\bibfield
		{journal} {\bibinfo  {journal} {Physical Review Letters}\ }\textbf {\bibinfo
			{volume} {99}},\ \bibinfo {pages} {046602} (\bibinfo {year}
		{2007})}\BibitemShut {NoStop}%
\end{thebibliography}

\begin{thebibliography}{10}
	
	\bibitem{Warot2004}
	B~Warot, J~Imrie, a.K Petford-Long, J.H Nickel, and T.C Anthony.
	\newblock {Influence of seed layers on the microstructure of NiFe layers}.
	\newblock {\em Journal of Magnetism and Magnetic Materials},
	272-276:E1495--E1496, May 2004.
	
	\bibitem{Gong2000}
	H.~Gong, D.~Litvinov, T.J. Klemmer, D.N. Lambeth, and J.K. Howard.
	\newblock {Seed layer effects on the magnetoresistive properties of NiFe
		films}.
	\newblock {\em IEEE Transactions on Magnetics}, 36(5):2963--2965, 2000.
	
	\bibitem{Jin2013}
	Lichuan Jin, Huaiwu Zhang, Xiaoli Tang, Feiming Bai, and Zhiyong Zhong.
	\newblock {Effects of ruthenium seed layer on the microstructure and spin
		dynamics of thin permalloy films}.
	\newblock {\em Journal of Applied Physics}, 113(5):053902, 2013.
	
	\bibitem{Yeh1987}
	T~Yeh, JM~Sivertsen, and JH~Judy.
	\newblock {Thickness dependence of the magnetoresistance effect in RF sputtered
		thin permalloy films}.
	\newblock {\em Magnetics, IEEE Transactions \ldots}, M(5):2215--2217, 1987.
	
	\bibitem{Rijks1995}
	TGSM Rijks and R~Coehoorn.
	\newblock {Semiclassical calculations of the anisotropic magnetoresistance of
		NiFe-based thin films, wires, and multilayers}.
	\newblock {\em Physical Review B}, 51(1), 1995.
	
	\bibitem{Choe1999}
	G.~Choe and M.~Steinback.
	\newblock {Surface roughness effects on magnetoresistive and magnetic
		properties of NiFe thin films}.
	\newblock {\em Journal of Applied Physics}, 85(8):5777, 1999.
	
	\bibitem{Acharyya2012}
	R~Acharyya.
	\newblock {\em {Spin dependent transport studies in magnetic, non-magnetic,
			antiferromagnetic, and half metals}}.
	\newblock PhD thesis, Michigan State University, 2012.
	
	\bibitem{Choi2002}
	Jongwan Choi, Youngmin Choi, Jongin Hong, Huyong Tian, Jae-Sung Roh, Younsoo
	Kim, Taek-Mo Chung, Young~Woo Oh, Yunsoo Kim, Chang~Gyun Kim, and Kwangsoo
	No.
	\newblock {Composition and Electrical Properties of Metallic Ru Thin Films
		Deposited Using Ru(C 6 H 6 )(C 6 H 8 ) Precursor}.
	\newblock {\em Japanese Journal of Applied Physics}, 41(Part 1, No.
	11B):6852--6856, November 2002.
	
	\bibitem{Mayadas1974}
	A.~F. Mayadas.
	\newblock {Resistivity of Permalloy thin films}.
	\newblock {\em Journal of Applied Physics}, 45(6):2780, 1974.
	
	\bibitem{Stiles1999}
	M.~Stiles and R.~McMichael.
	\newblock {Model for exchange bias in polycrystalline
		ferromagnet-antiferromagnet bilayers}.
	\newblock {\em Physical Review B}, 59(5):3722--3733, February 1999.
	
	\bibitem{Fang2011}
	D~Fang, H~Kurebayashi, J~Wunderlich, K~V\'{y}born\'{y}, L~P Z\^{a}rbo, R~P
	Campion, A~Casiraghi, B~L Gallagher, T~Jungwirth, and A~J Ferguson.
	\newblock {Spin-orbit-driven ferromagnetic resonance.}
	\newblock {\em Nature nanotechnology}, 6(7):413--7, July 2011.
	
	\bibitem{Liu2011}
	Luqiao Liu, Takahiro Moriyama, D.~Ralph, and R.~Buhrman.
	\newblock {Spin-Torque Ferromagnetic Resonance Induced by the Spin Hall
		Effect}.
	\newblock {\em Physical Review Letters}, 106(3):1--4, January 2011.
	
\end{thebibliography}
\end{document}